\documentclass[preprint]{elsarticle}

\usepackage{hyperref}
\usepackage{mathtools}
\usepackage{amsmath}
\usepackage{float}
\usepackage{subcaption} 
\usepackage{amssymb}
\usepackage{todonotes}
\usepackage{sidecap}

\makeatletter
\def\ps@pprintTitle{%
   \let\@oddhead\@empty
   \let\@evenhead\@empty
   \let\@oddfoot\@empty
   \let\@evenfoot\@oddfoot
}
\makeatother

\begin{document}

\begin{frontmatter}

\title{A mathematical model of cell fate selection on a dynamic tissue}

\author[1]{Domenic P.J. Germano} \ead{germanod@student.unimelb.edu.au}
\author[1]{James M. Osborne\corref{cor1}} \ead{jmosborne@unimelb.edu.au}

\cortext[cor1]{Corresponding author}

\address[1]{School of Mathematics and Statistics, The University of Melbourne, Parkville, Victoria 3010, Australia}

\begin{abstract}
Multicellular tissues are the building blocks of many biological systems and organs. 
These tissues are not static, but dynamically change over time. 
Even if the overall structure remains the same there is a turnover of cells within the tissue.
This dynamic homeostasis is maintained by numerous governing mechanisms which are finely tuned in such a way that the tissue remains in a homeostatic state, even across large timescales. 
Some of these governing mechanisms include cell motion, and cell fate selection through inter cellular signalling. 
However, it is not yet clear how to link these two processes, or how they may affect one another across the tissue. 
In this paper, we present a multicellular, multiscale model, which brings together the two phenomena of cell motility, and inter cellular signalling, to describe cell fate selection on a dynamic tissue. 
We find that the affinity for cellular signalling to occur greatly influences a cells ability to differentiate. 
We also find that our results support claims that cell differentiation is a finely tuned process within dynamic tissues at homeostasis, with excessive cell turnover rates leading to unhealthy (undifferentiated and unpatterned) tissues.
\end{abstract}


\end{frontmatter}

\section{Introduction}
 are comprised of dynamic multicellular tissues, with many healthy tissues existing in a state of dynamics homeostasis. This homeostasis is achieved through a balancing of multiple interacting governing mechanisms, including (but not limited to) the coordination of cell differentiation and proliferation. The process of cell differentiation is  known to be mediated by the Notch pathway, by one of three main mechanisms: lateral inhibition, lateral induction or lineage decisions \cite{Borggrefe2018}. The Notch pathway controls cell differentiation in many tissues, including muscle, nervous system and intestine \cite{Koch689}. Bar a few simple pedagogical models  \cite{osborne2017comparing, toth2017}, how the two processes of differentiation and proliferation interact at different scales is not well studied. The intestinal epithelium is one of the fastest self renewing tissues in the human body, meaning the interaction of cell differentiation and proliferation is of particular interest. Tissue renewal is controlled within millions of test-tube-like structures which line the intestinal walls, known as the crypts of Lieberkühn \cite{shanmugathasan2000apoptosis}.
At the base of each crypt, resides a stem cell population, which is directly responsible for the replenishment of the epithelial cells through mitosis.
After dividing, cells migrate up the walls of the crypt, differentiating as they go, where upon reaching the top of the crypt they form the epithelial lining of the intestine and are removed by sloughing to prevent overcrowding.
In humans, this renewal process usually takes 6--7 days \cite{shanmugathasan2000apoptosis}.
Therefore, healthy crypt homeostasis is controlled by the fine balance of cell proliferation, migration, differentiation, and apoptosis \cite{shanmugathasan2000apoptosis}. 
As cells migrate up the crypt wall, they differentiate into a committed cell fate. This process is controlled by Delta-Notch signalling, via lateral inhibition, the process where a cell prevents adjacent cells from adopting the same cell fate \cite{collier1996pattern}.

The key mechanisms controlling Delta-Notch signalling are the Notch receptors and Delta ligands, both of which are transmembrane proteins.  The pathway is activated when Delta ligands on one cell bind to Notch receptors of a neighbouring cell. 
As a response, intracellular reactions are triggered, releasing Notch intracellular domain, on the adjacent cell, which in turn allows target gene expression, and thus leads to cell differentiation \cite{bray2006notch}.

An understanding as to how crypt dynamics, and more generally tissue dynamics, are maintained at homeostasis, and what may lead to deviations from such homeostasis is highly desirable. Mathematical and computational models provide an ideal framework to study these dynamical tissues. 
Previous modelling work has provided insight into how Delta-Notch signalling occurs within  dynamic tissues.
One early model of Delta-Notch signalling was proposed by \citeauthor{collier1996pattern} \cite{collier1996pattern}, where a tissues of static cells interact with their nearest neighbours through lateral inhibition. The authors found that their model reproduced Notch patterning, analogous to those found in living systems. 
In 2011 \citeauthor{buske2011comprehensive} \cite{buske2011comprehensive} coupled a logic based Delta-Notch signalling with an over-lapping spheres model of cell dynamics with position dependent proliferation. Using their model, the authors were able to reproduce the correct ratios of cell types present at steady state. The model can also describe and predict dynamic behaviour of the epithelium at both steady state, and also following the introduction of mutant cells. The authors also showed that the intestinal epithelium is capable of complete recovery following eliminations of each subpopulations of cell within the crypt epithelium.
The \citeauthor{collier1996pattern} model for Delta Notch has been coupled with numerous multicellular models. \citeauthor{osborne2017comparing} \cite{osborne2017comparing} present a comparison of coupling the signaling model with five biomechanical models and show that as long as the use of biomechanical model is appropriate (i.e there are no model artefacts), then the biomechanical models are equivalent.
The \citeauthor{collier1996pattern} model was extended to include other aspects of signalling in the Crypt in \citeauthor{kay2017role} \cite{kay2017role}. In the paper the authors analyse this mode on pairs of connected cells and show that the Delta Notch patterning can influence cell differentiation.

Furthermore, it is known that during development, cell morphology changes, and therefore the contact geometry between neighbouring cells changes too. Since Notch signalling is mediated via transmembrane proteins, these morphological changes could influence cellular communication. To investigate, \citeauthor{shaya2017cell} developed a model of Notch signalling which is dependent upon the contact area between neighbouring cells \cite{shaya2017cell}. They found that contact area biases cellular differentiation, where smaller cells are more likely to differentiate into the primary cell fate.

It is clear that understanding the interplay between cell fate signalling and cell dynamics is crucial in furthering our knowledge of tissue and organ development and function. 
To the authors knowledge, no previous study has investigated how properties of the inter-cellular signalling, and cell turnover, influences cell differentiation and fate selection. 
In this paper we present a multiscale, multicellular model which couples cell dynamics with Delta-Notch signalling and cell fate selection. Despite motivation for this work being the colonic crypt, where tissue self renewal occurs at a relatively fast timescale, this model is not specific to a particular tissue. The tissue geometry and cell cycle duration can be calibrated to study tissues of interest.

The remainder of this paper is structured as follows, we begin by presenting our multicellular multiscale model of cell fate selection in a dynamic tissue in Section \ref{sec:model}.
In Section \ref{sec:results}, we first present an investigation into how subcellular dynamics and tissue geometry influences pattern formation on static tissues, before demonstrating the influence of cell turnover on patterning. 
Finally in Section \ref{sec:conclusion}, we discuss our results and relate them back to biological dynamical tissues like the colorectal crypt.

\section{Model}
\label{sec:model}
Here, we model the tissue as a collection of discrete, interacting, individual cells, more commonly referred to as a multicellular model. In a multicellular model, cells are represented as a single point (or a collection of points) in space, and allows details at the cellular level and tissue level to be included \citep{osborne2017comparing}. 
Specifically, they allow the inclusion of cell population turnover and cellular signalling, which occur at differing timescales.%, through a framework known as multiscale modelling.
Below, we discuss the multicellular, multiscale model we use to couple these processes, in order to analyse cell fate selection within a dynamic tissue.

\subsection{Biomechanical Model}
To describe cell dynamics, we use a lattice-free, cell-centred model \cite{meineke2001cell}. 
The net force on a given cell $i$, $\mathbf{F}_{i}^{\text{Net}}$, is found by balancing the force due to neighbouring cell interactions, $\mathbf{F}_{i}^{\text{Interactions}}$, and the viscous forces on the cell, $\mathbf{F}_{i}^{\text{Viscous}}$, as proposed by \citeauthor{meineke2001cell} \citep{meineke2001cell}:
\begin{align}
\mathbf{F}_{i}^{\text{Net}} =\mathbf{F}_{i}^{\text{Interactions}} +  \mathbf{F}_{i}^{\text{Viscous}}, \qquad \forall i.
\end{align}
Due to the highly viscous environment that cells occupy, we assume that cell motion is over-damped \citep{dallon2004cellular}, and therefore viscous forces dominate allowing us to neglect inertial terms and all motion is determined by a force balance, $\mathbf{F}_{i}^{\text{Net}}=\mathbf{0}$.

\subsubsection{Equations of Motion}
We follow \citeauthor{meineke2001cell} \citep{meineke2001cell} and model the force due to neighbouring cell interactions using a linear Hooke's law acting at cell-centres to describe attraction and repulsion between neighbouring cells.
We also assume viscous forces acting on the cell oppose the direction of motion. 
This leads to the equations of motions, for cell $i$, at position $\mathbf{r}_i$:
\begin{align}
\nu \frac{d \mathbf{r}_i}{dt}  = \sum_{j\in M_i} k_{ij}^{sp} \left( \vert \mathbf{r}_{ij}\vert - s_{ij} \right) \hat{ \mathbf{r}}_{ij}, \qquad \forall i,
\end{align}
where $\mathbf{r}_{ij} = \mathbf{r}_{j}-\mathbf{r}_{i}$ is the displacement between cells $i$ and $j$, $\nu$ is the drag coefficient applied to all cell centres, $M_i$ the neighbouring cells of cell $i$, $k_{ij}^{sp}$ the spring constant between cells $i$ and $j$ (here taken to be constant so  \mbox{$k_{ij}^{sp}=k^{sp}$}), and  $s_{ij} = s_{ij}(t)$ is the length separation between cells $i$ and $j$, which has the form:
\begin{align}
s_{ij}(t) = \begin{cases}
\varepsilon + \tau_i(1-\varepsilon) , &\tau_i \le 1,\\
1, &\text{otherwise},
\end{cases} \label{Eq_Sij}
\end{align}
with $\varepsilon >0$ being the initial separation between daughter cells, and $\tau_i \ge 0$ the age of cell $i$ at time $t$. All parameters are given in Table~\ref{table:parameter_values}. Note all distances are measured in cell diameters (cd) which we take to be the average size of a crypt epithelial cell 10$\mu m$ \cite{dallon2004cellular}.%, REF\todo{add crypt ref for epithelial cell size}.  

\subsubsection{Cell Population Turnover}
Our aim is to get a cell turnover rate of $\gamma$ divisions (and deaths) per hour per cell. In order to do this we model each cell to have its own cell cycle duration, $T$, which is sampled from a Uniform distribution\footnote{similar results can be obtained by sampling from a Gamma distribution of $ T \sim \mathbf{ \Gamma}\left( 48, 48 \gamma \right)$.}:
\begin{align}
T \sim \mathbf{U}\left(\frac{3}{4\gamma} , \frac{5}{4\gamma} \right),
\end{align}
where $\gamma$ is the target cell turnover rate. When a cell reaches its cell cycle duration, it is labelled the parent cell, with position $\mathbf{r}^p$, and subsequently proliferates into two daughter cells, with positions $\mathbf{r}_i$ and $\mathbf{r}_j$, displaced at a separation $\varepsilon>0$ apart, along a randomly directed unit normal $\hat{\mathbf{n}}$:
\begin{align}
\mathbf{r}_i &= \mathbf{r}^p + \frac{\varepsilon}{2} \hat{\mathbf{n}}, \qquad \mathbf{r}_j = \mathbf{r}^p - \frac{\varepsilon}{2} \hat{\mathbf{n}}.
\end{align}
To represent a turnover of cells, and to maintain a fixed number of cells in the tissue, whenever a cell divides we randomly select a cell to be removed in the same timestep.

\subsection{Biochemical Model}
The model of intracellular signalling we employ was initially described by \citeauthor{collier1996pattern}  \cite{collier1996pattern}. \citeauthor{collier1996pattern} consider a simplified model of a cell, focusing on the Delta-Notch pathway, where inhibited cells have a reduced ability to inhibit other cells. The key mechanisms of the Delta-Notch pathway are the Notch receptors and Delta ligands, both of which are transmembrane proteins. The pathway is activated when Delta ligands on one cell bind to Notch receptors of neighbouring cells. As a response, intracellular reactions are triggered, which in turn allows target gene expression, and thus leads to cell differentiation \cite{bray2006notch}. The biochemical model assumptions are summarised by the following:
\begin{enumerate}
\item Only cells in direct contact may interact via Delta-Notch signalling.
\item The rate of Notch production is an increasing function of the amount of Delta present in neighbouring cells.
\item The rate of Delta production is a decreasing function of the amount of Notch within the same cell.
\item The rate of both Notch and Delta decay obey exponential laws, with rates $\mu$ and $\rho$ respectively.
\item A cell's fate is determined by the amount of Notch within the cell.
\end{enumerate}
These model assumptions give rise to the following (non-dimensional) mathematical model, for cell $i$, with Notch level $N_i$ and Delta level $D_i$:
\begin{align}
\frac{d N_i}{dt} &=  \mu \frac{{\bar{D}}^k_i}{a + {\bar{D}}^k_i}  - \mu N_i ,\qquad \forall i,\\
\frac{d D_i}{dt} &=  \rho \frac{1}{1 + b {N}_i^h} - \rho D_i , \qquad \forall i.
\end{align}
where $a$ is the Notch affinity constant (which dictates how readily Delta ligands bind to Notch receptors), and $b$ the Delta affinity constant (which dictates how readily Delta ligands are produced), and $k$ and $h$ the exponent for Notch and Delta synthesis respectfully. Lastly, $\bar{D}_i$ is the mean level of Delta within neighbouring cells (of cell $i$).
On cell division the level of Delta and Notch in the daughter cells are assigned to be the same as the parent cell.

\subsection{Tissue Geometry and Initial Conditions}
\begin{figure}[H]
\centering
\includegraphics[width = 0.5\textwidth]{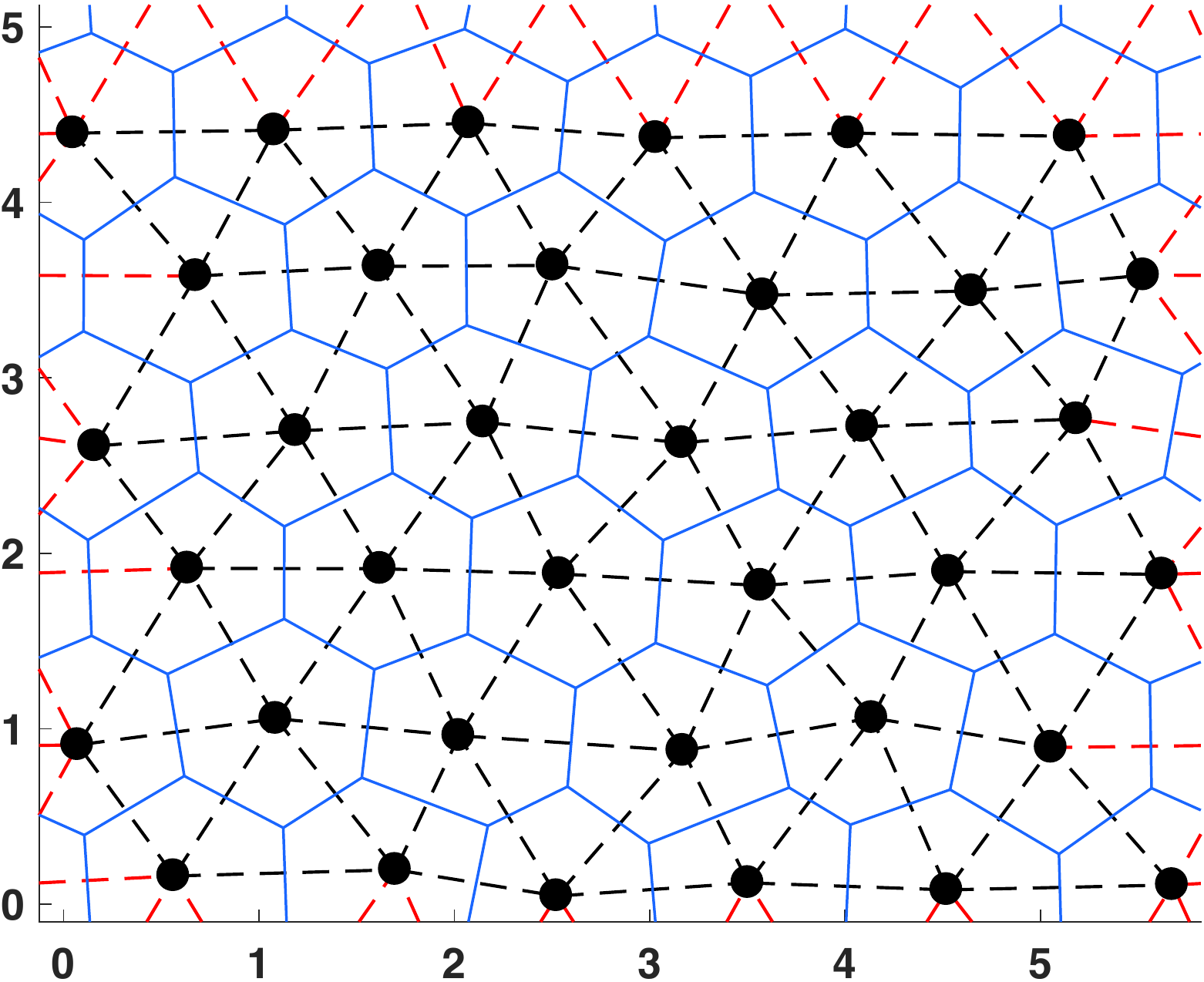}
\caption{\label{tissue_struct} A typical tissue geometry. The black dots show cell centres. The dashed lines show the neighbouring cells defined via a Delaunay Triangulation, with black dashes showing immediate neighbouring cells and red dashes describing neighbours due to periodicity. The solid blue lines describe the cell shapes, given by a Voronoi tesselation.}
\end{figure}

We constrain the tissue to lie in a domain of size $L_{x} \times L_{y}$, where \mbox{$L_{x},L_{y} \in \mathbb{R}$}, with horizontally and vertically periodic boundaries (a toroidal domain). Due to the honeycomb structure of centre based cells in equilibrium it is more informative to describe the tissue size by referring to the number of horizontally and vertically stacked cells, $C_{x} \times C_{y}$, where \mbox{$C_{x},C_{y} \in \mathbb{N}$}. The relation between the number of cells horizontally stacked ($C_{x}$), and the horizontal length ($L_{x}$) is \mbox{$L_{x} = C_{x}$}. However, because we initialise the tissue on a honeycomb, hexagonal lattice, which is the equilibrium state for our biomechanical model \cite{thompson1942growth}, the relation between the number of cells vertically stacked ($C_{y}$), and the vertical length ($L_{y}$) is \mbox{$L_{y} = \frac{\sqrt{3}}{2} C_{y}$}. We treat each cell as a distinct agent,  which interacts with it's neighbouring cells. These neighbouring cells are determined by a Delaunay Triangulation between cell centres. The shape of each cell is given by a Voronoi tessellation, which is the natural dual of the Delaunay Triangulation. Figure \ref{tissue_struct} shows a typical tissue geometry with $L_{x}=6$cd and $L_{y}=3\sqrt{3}$cd ($C_{x}=C_{y}=6$).
%We constrain the tissue to lie in a domain of size $L_{x} \times L_{y}$, where \mbox{$L_{x},L_{y} \in \mathbb{N}$}, (with $L_{x}=10$cd by $L_{y}=10$cd unless otherwise stated)\todo{Check this or introduce variables for width and height} with horizontally and vertically periodic boundaries (a toroidal domain). Cells are initialised on a hexagonal array, which is the equilibrium state for our biomechanical model \cite{thompson1942growth}. We treat each cell as a distinct agent,  which interacts with it's neighbouring cells. These neighbouring cells are determined by a Delaunay Triangulation between cell centres. The shape of each cell is given by a Voronoi tessellation, which is the natural dual of the Delaunay Triangulation. Figure \ref{tissue_struct} shows a typical tissue geometry.

Unless stated otherwise, the initial conditions we use for the biochemical model are uniform homogeneous. That is, for cell $i$, the initial levels of Notch and Delta within the cell are:
\begin{align}
N_i(0) = N^0, \quad D_i(0) = D^0, \quad \forall i,
\end{align}
where $0 \leq N^0 \leq 1$ and $0 \leq D^0 \leq 1$.

\section{Results}
\label{sec:results}
We first consider the behaviour of a static tissue (i.e. cells have no turnover rate). Therefore, cells remain stationary and only interact with a fixed set of six neighbouring cells. In Section~\ref{sec:dynamic} we relax this assumption and allow cells to proliferate, undergo apoptosis and move.

\subsection{Tissue Geometry and Biochemical Initial Conditions Support Patterning}

\begin{figure}[htbp]
%\captionsetup[subfigure]{slc=off,margin={0cm,1cm,-1cm}}
\captionsetup[subfigure]{slc=off,skip=-0.25cm,margin={0.3cm,0cm,0cm}}
\centering
\begin{subfigure}{.08\textwidth}
  \centering \rotatebox{90}{\textbf{Homogeneous}}
\end{subfigure}
\begin{subfigure}{.45\textwidth}
  \centering
      \textbf{Steady State Notch}\par\medskip
  \caption{\label{fig:pat_hom}}
  \includegraphics[height= 0.13\textheight]{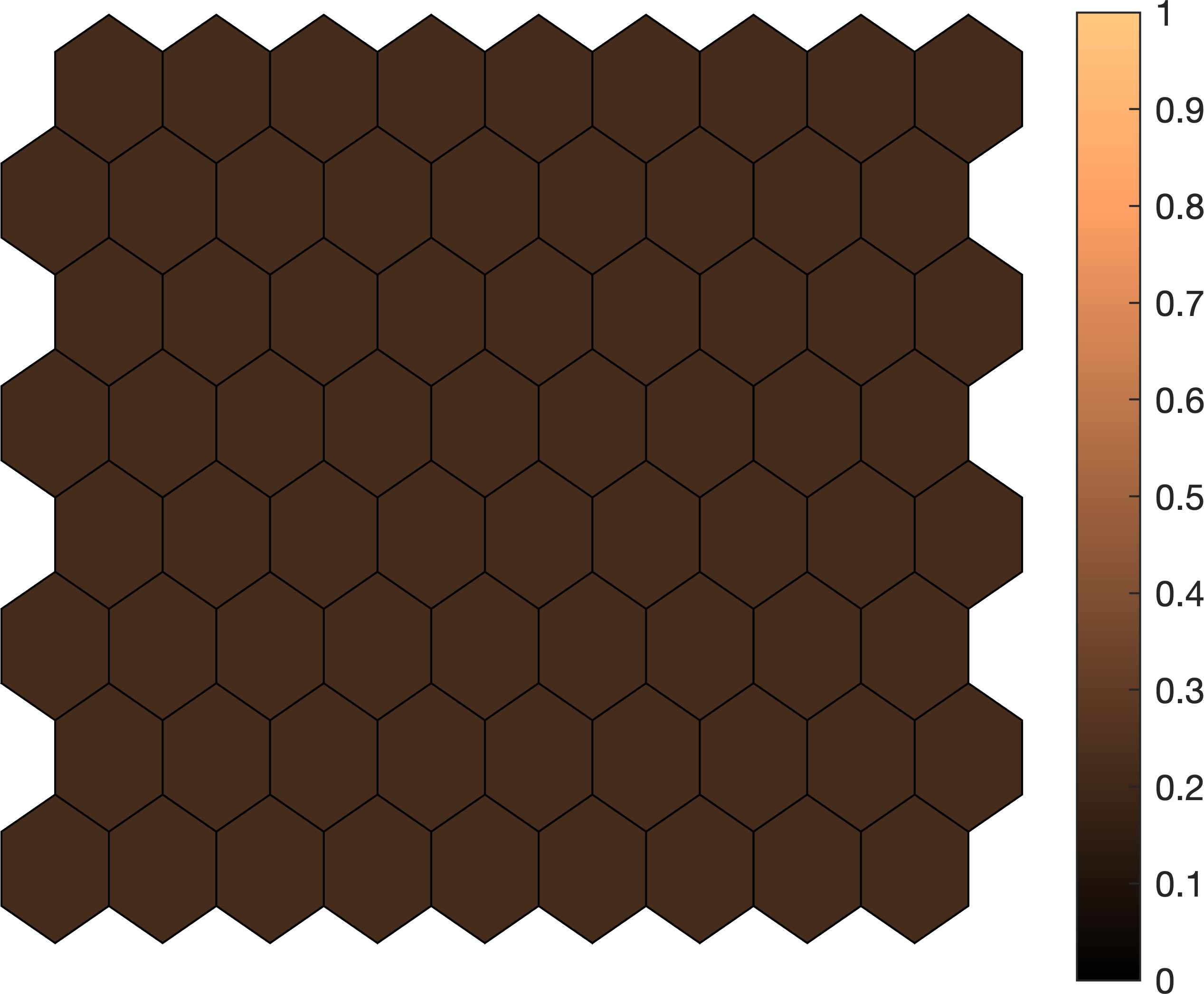}
\end{subfigure}
\begin{subfigure}{.45\textwidth}
  \centering
        \textbf{Intracellular Notch Solution}\par\medskip
    \caption{\label{fig:notch_hom}}
  \includegraphics[height=0.13\textheight]{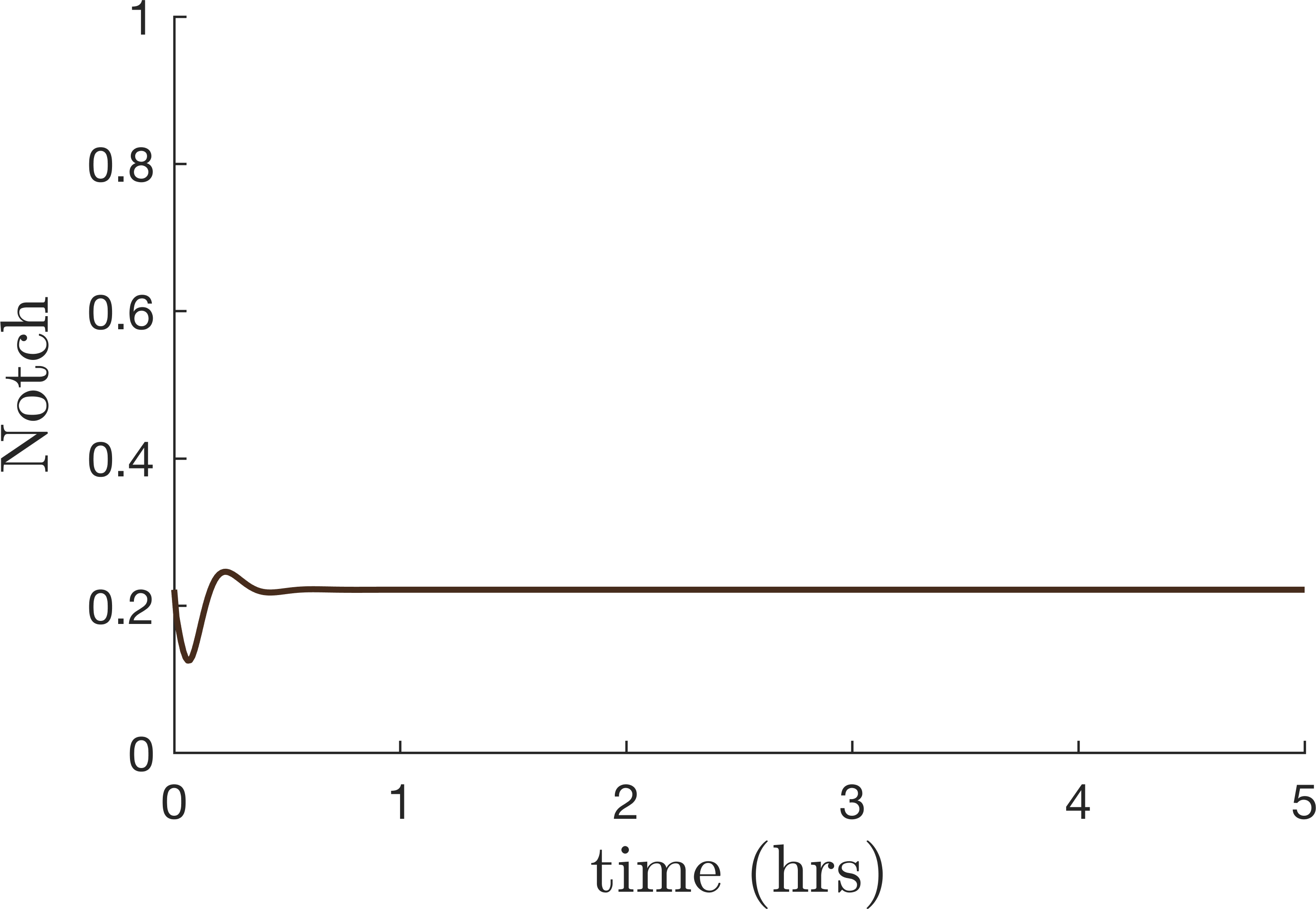}
\end{subfigure}

  \begin{subfigure}{.08\textwidth}
  \centering {\rotatebox{90}{\textbf{Single Seed}}}
\end{subfigure}
\begin{subfigure}{.45\textwidth}
  \centering
    \caption{\label{fig:pat_9b8}}
    \includegraphics[height=0.13\textheight]{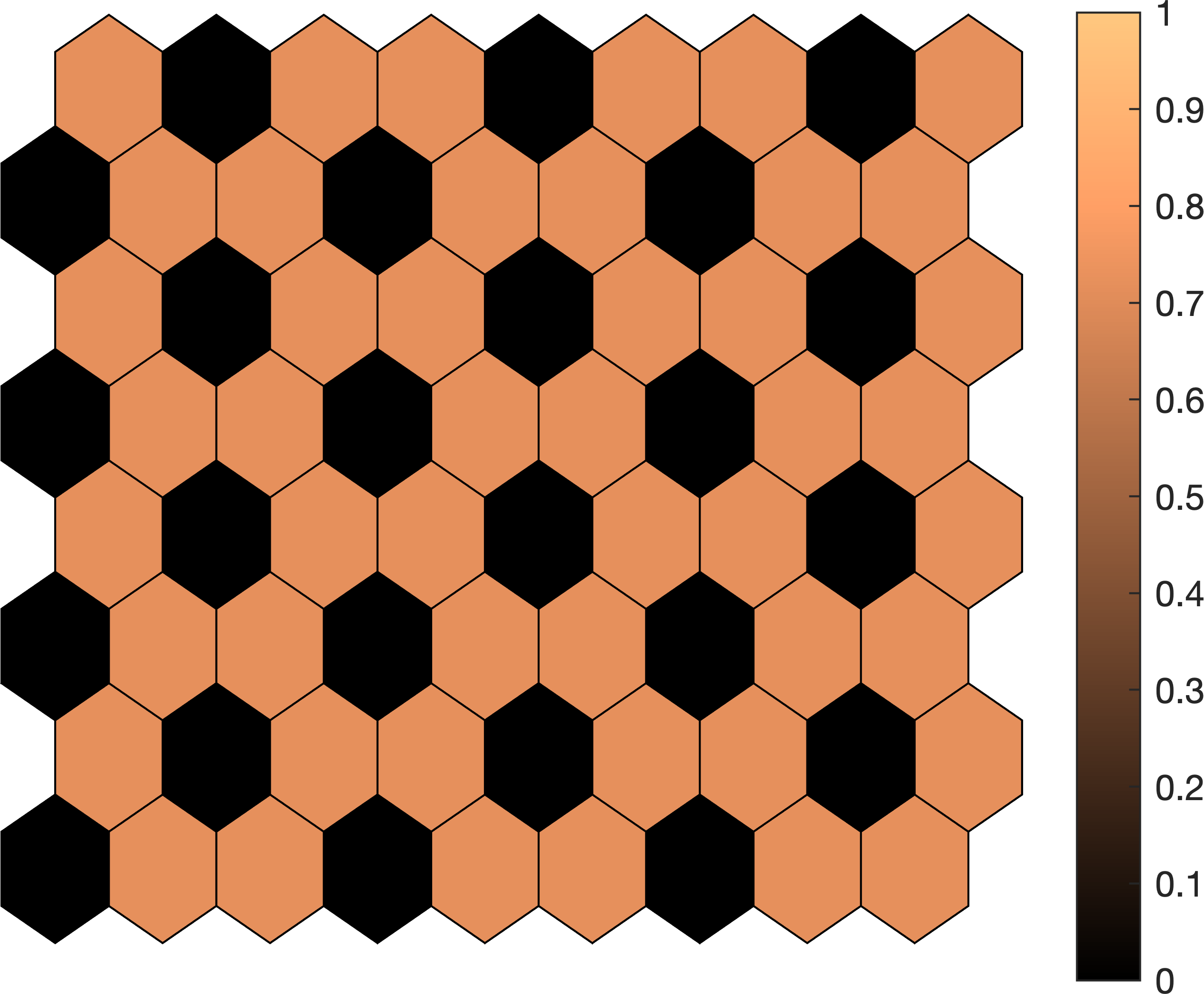}
\end{subfigure}
\begin{subfigure}{.45\textwidth}
  \centering
    \caption{\label{fig:notch_9b8}}
  \includegraphics[height=0.13\textheight]{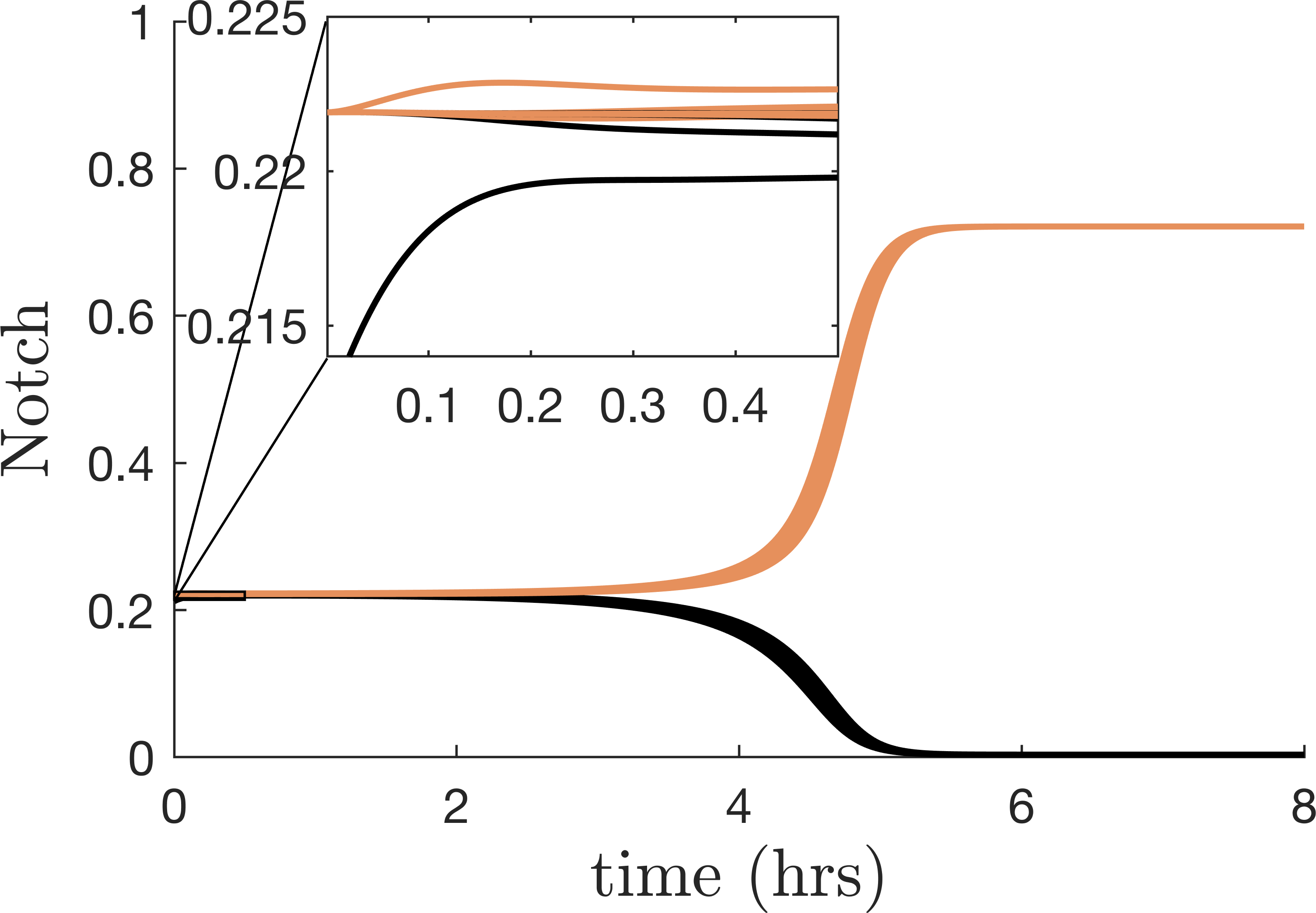}
\end{subfigure}

  \begin{subfigure}{.08\textwidth}
  \centering {\rotatebox{90}{\textbf{Random}}}
\end{subfigure}
\begin{subfigure}{.45\textwidth}
  \centering
    \caption{\label{fig:pat_rand_p}}
  \includegraphics[height=0.13\textheight]{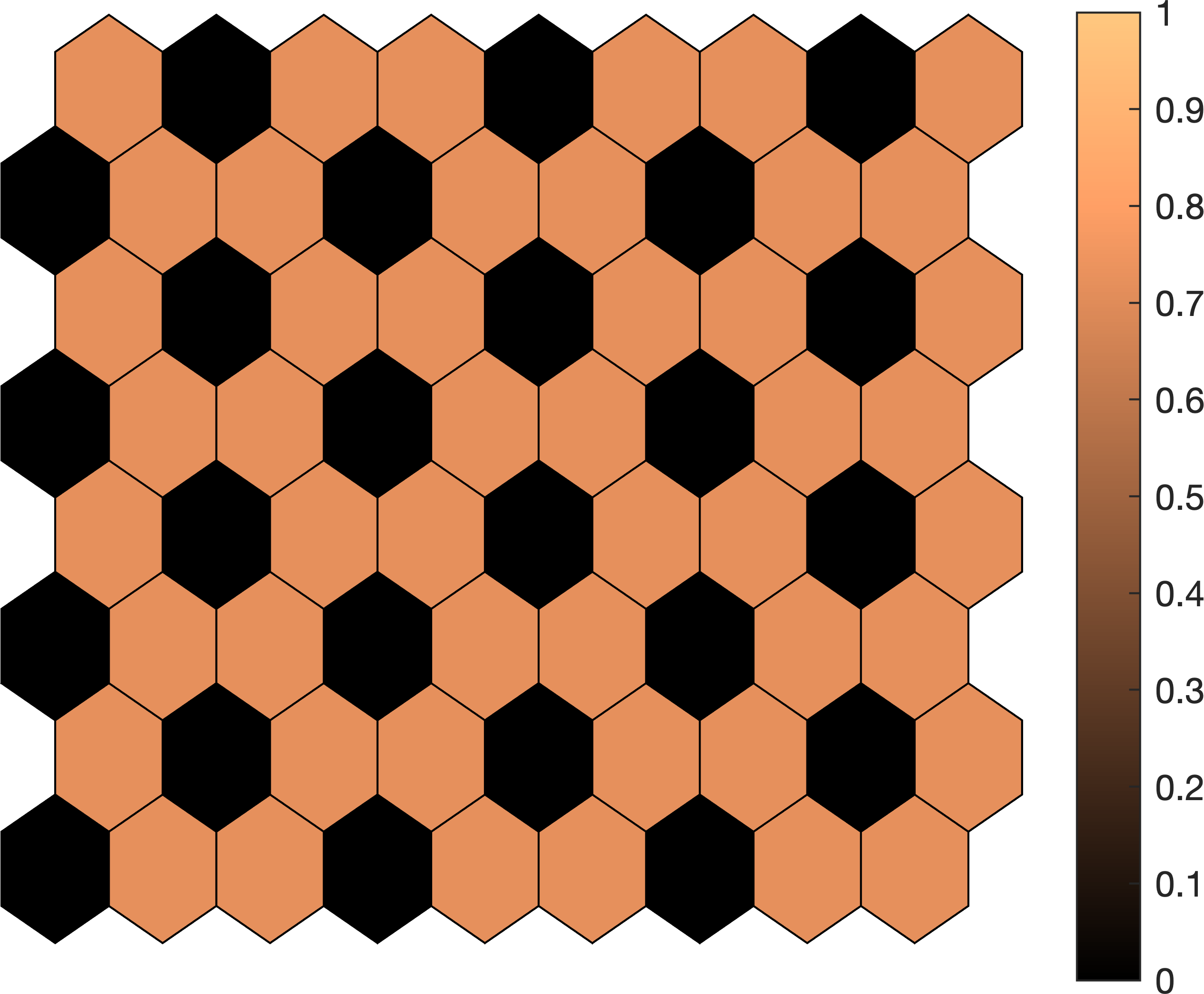}
\end{subfigure}
\begin{subfigure}{.45\textwidth}
  \centering
    \caption{\label{fig:notch_rand_p}}
  \includegraphics[height=0.13\textheight]{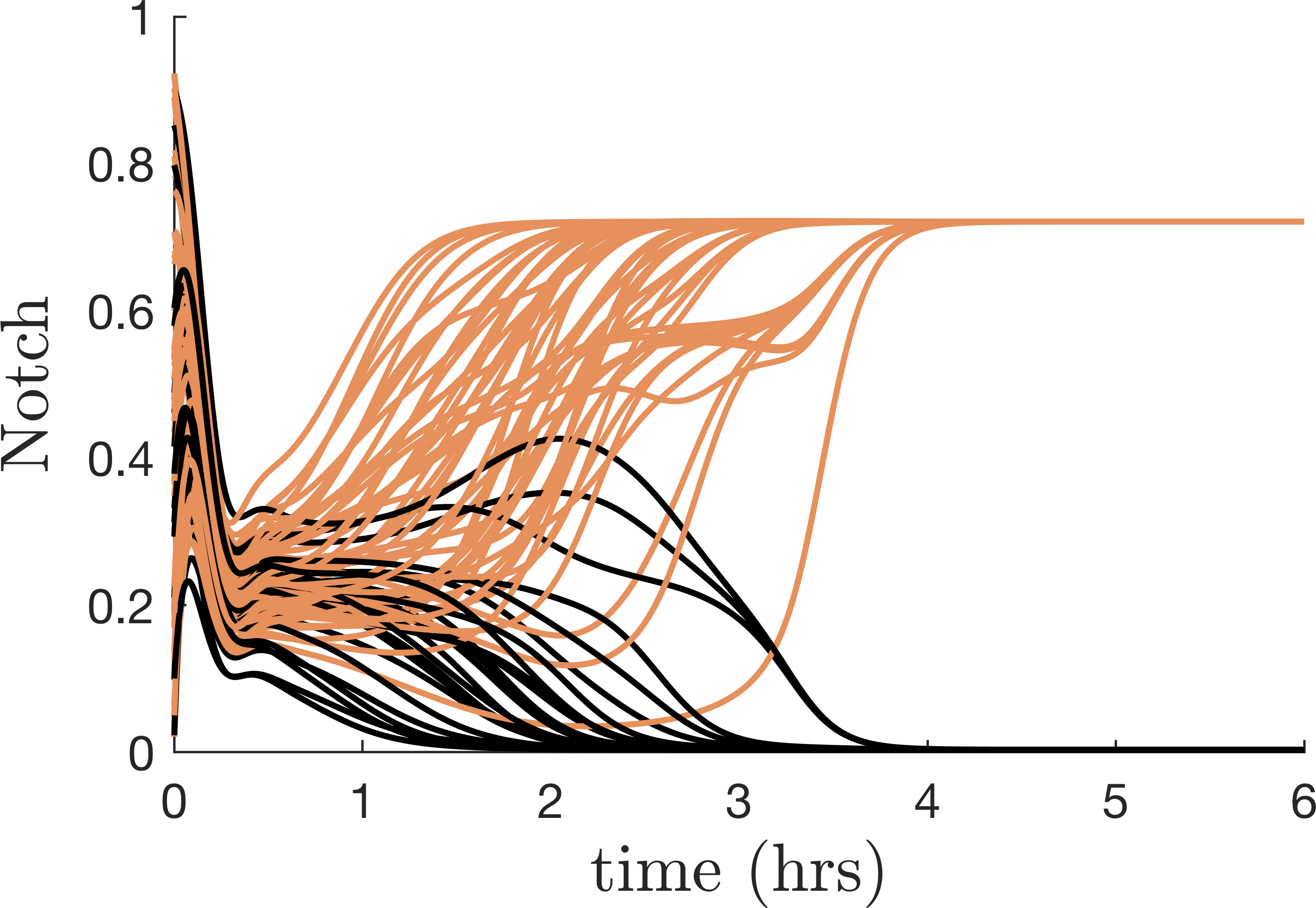}
\end{subfigure}

 \begin{subfigure}{.08\textwidth}
  \centering \rotatebox{90}{\textbf{Random}}
\end{subfigure}
\begin{subfigure}{.45\textwidth}
  \centering
    \caption{\label{fig:pat_rand}}
  \includegraphics[height=0.13\textheight]{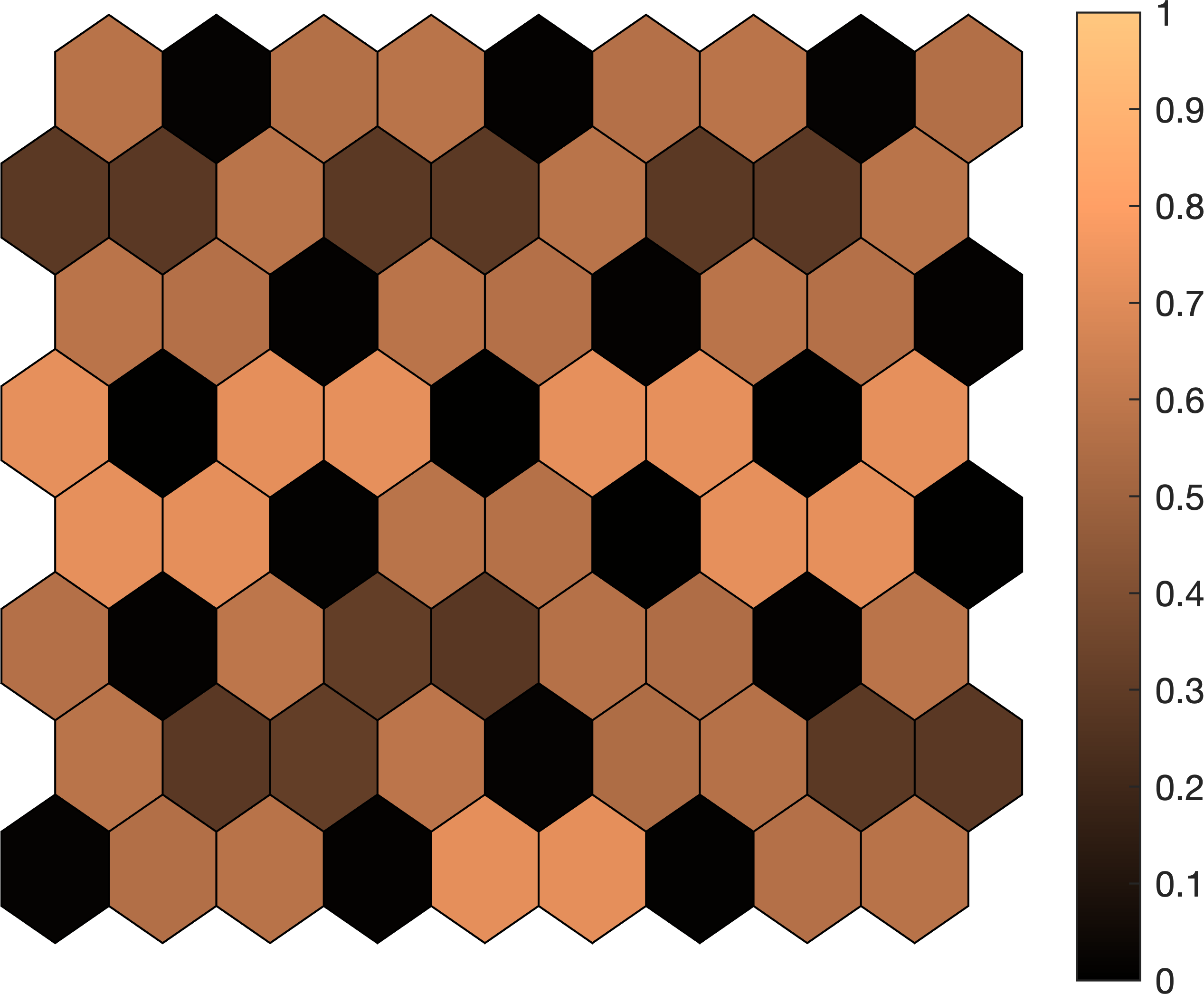}
\end{subfigure}
\begin{subfigure}{.45\textwidth}
  \centering
    \caption{\label{fig:notch_rand}}
 \includegraphics[height=0.13\textheight]{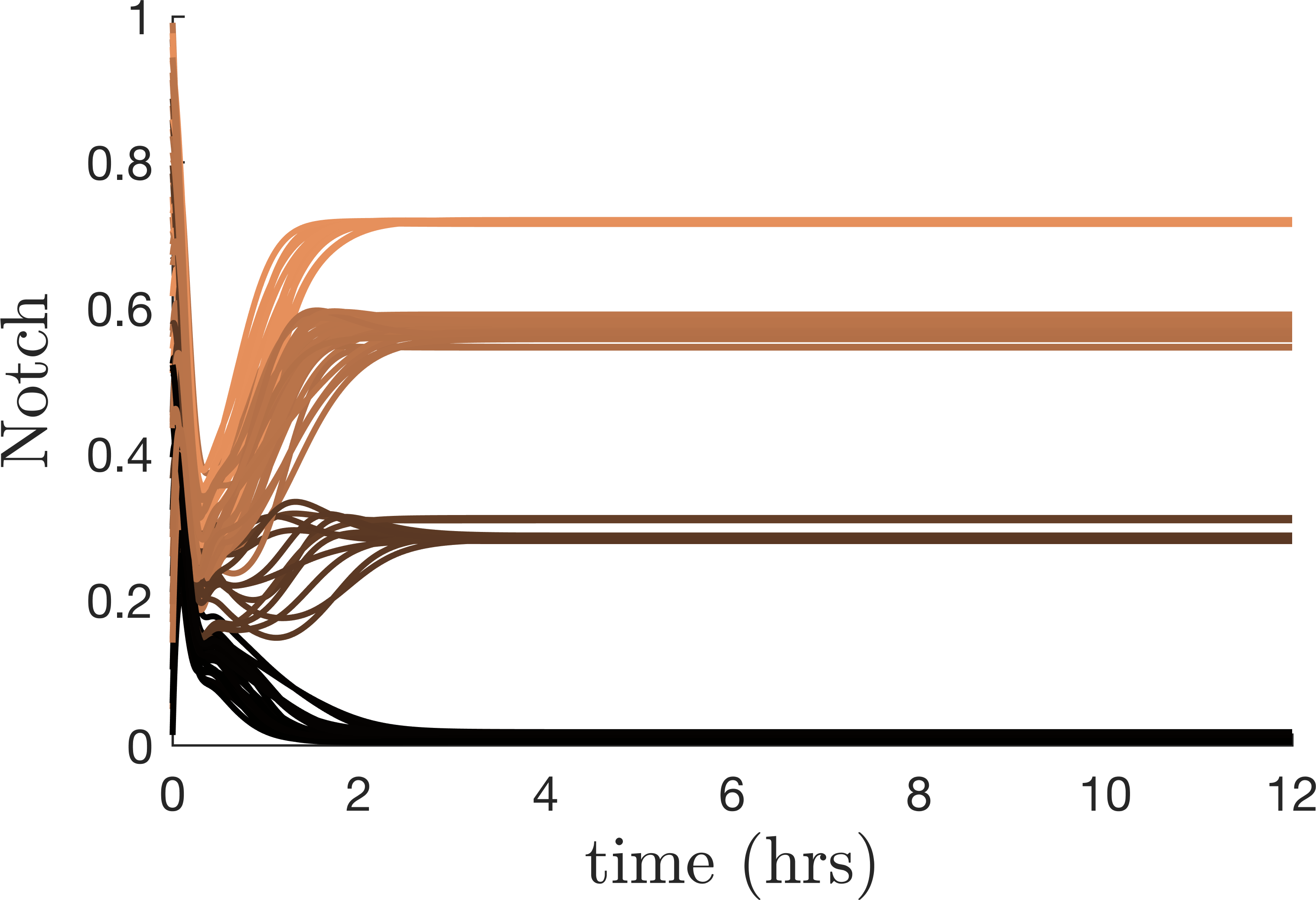}
\end{subfigure}

 \begin{subfigure}{.08\textwidth}
  \centering \rotatebox{90}{\textbf{Unfit Domain}}
\end{subfigure}
\begin{subfigure}{.45\textwidth}
  \centering
    \caption{\label{fig:10b8_p}}
  \includegraphics[height=0.13\textheight]{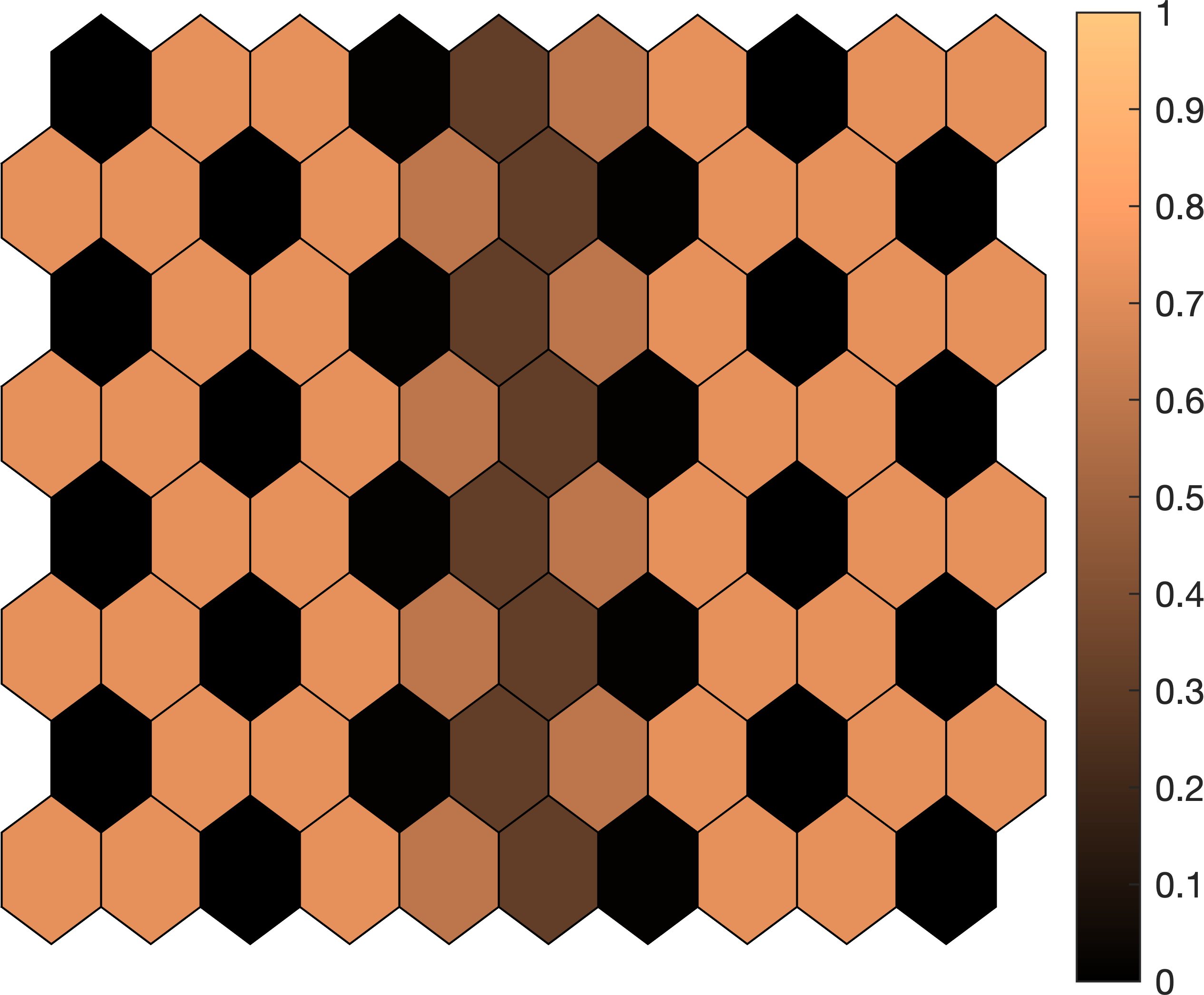}
\end{subfigure}
\begin{subfigure}{.45\textwidth}
  \centering
    \caption{\label{fig:10b8_n}}
 \includegraphics[height=0.13\textheight]{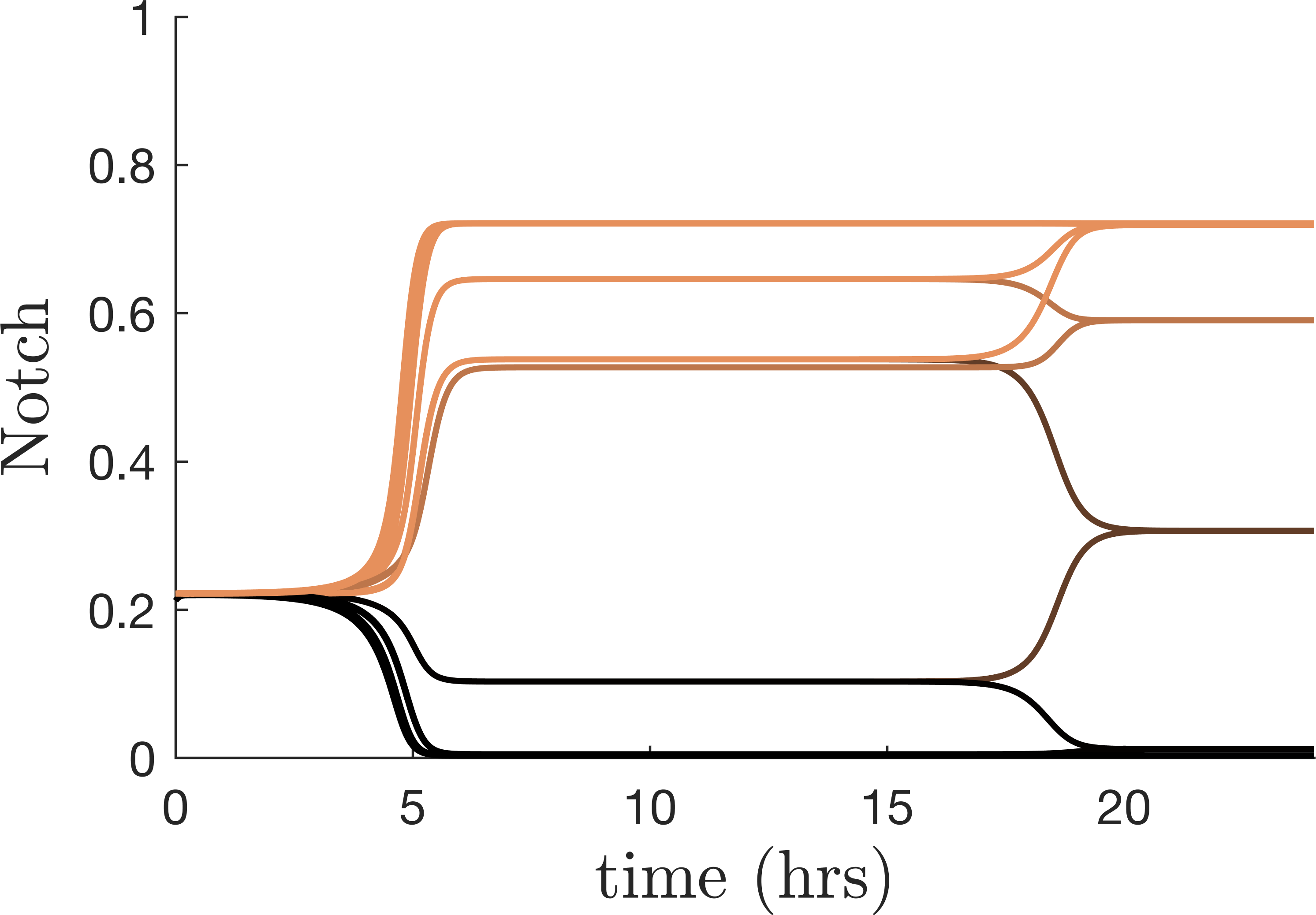}
\end{subfigure}

\vspace{0.5cm}
%\captionsetup[subfigure]{slc=off,skip=0.25cm,margin={1cm,0cm,0cm}}
\begin{subfigure}[b]{0.08\textwidth}
\hspace{\textwidth}
\end{subfigure}
\begin{subfigure}[b]{0.44\textwidth}
\caption{\label{fig:primary_cell}}
%\centering\includegraphics[height=0.035\textheight]{primary_cel}
\centering\includegraphics[height=0.035\textheight]{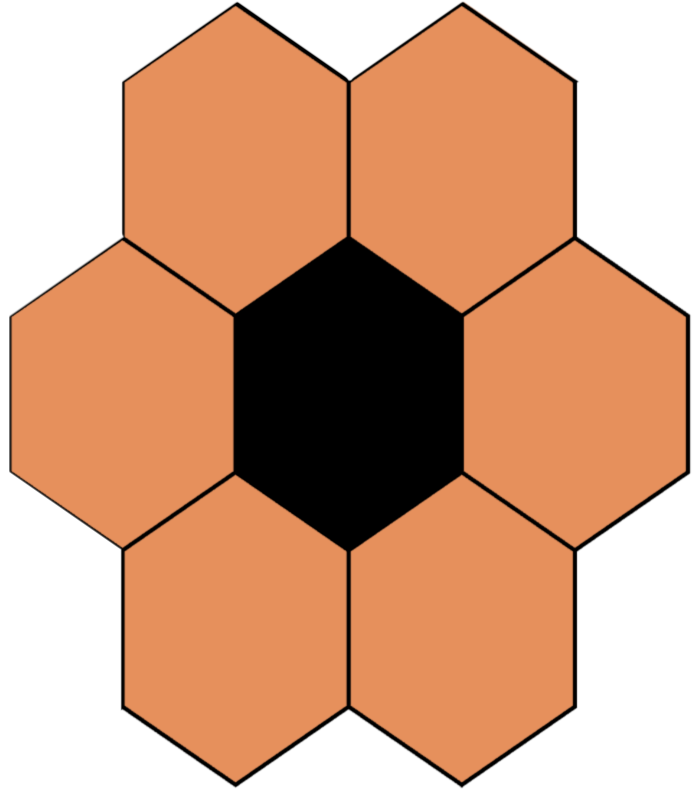}
\end{subfigure}
\begin{subfigure}[b]{0.44\textwidth}
\caption{\label{fig:secondary_cell}}
%\centering\includegraphics[height=0.035\textheight]{secondary_cel}
\centering\includegraphics[height=0.035\textheight]{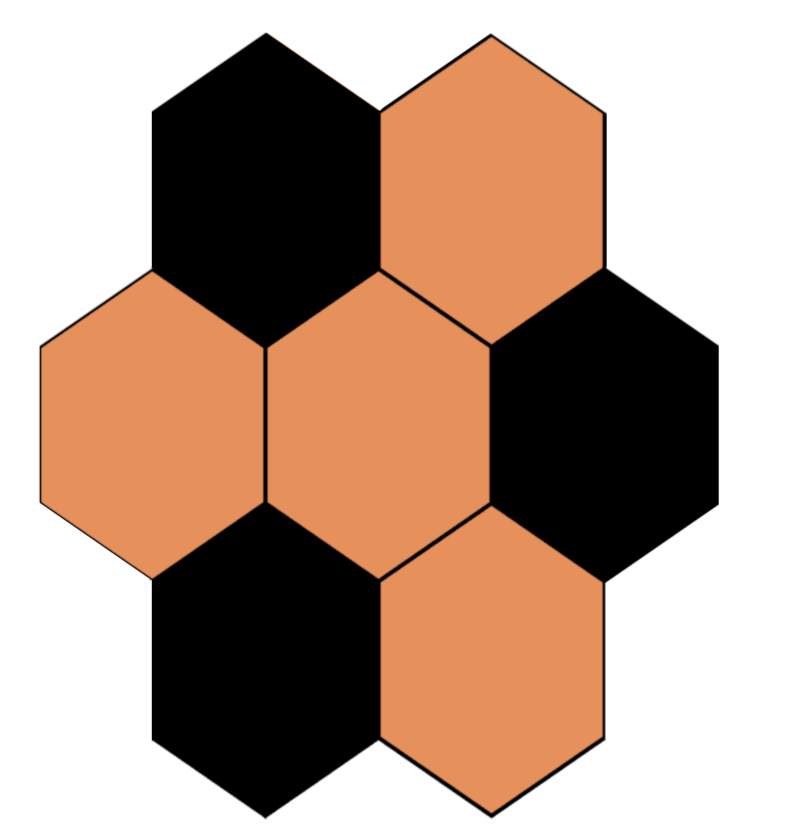}
\end{subfigure}
\caption{  \label{fig:PATTERN}  \textit{In silico} experiments with parameter values $a=0.1$, $b=100$, $\mu=\rho=10$, $k=h=2$, $\gamma\rightarrow\infty$, $\nu=1$ and $k^{sp}=50$. \subref{fig:pat_hom}, \subref{fig:pat_9b8}, \subref{fig:pat_rand_p}, \subref{fig:pat_rand} and \subref{fig:10b8_p} show the steady state Notch patterning for differing initial conditions of Notch. \subref{fig:notch_hom}, \subref{fig:notch_9b8}, \subref{fig:notch_rand_p}, \subref{fig:notch_rand} and \subref{fig:10b8_n} show the associated time series solutions of intracellular Notch. Note \subref{fig:notch_9b8} includes inset of early time solutions to show initial heterogeneity. Videos of these simulations can be found in SI Movie 1 -- SI Movie 5.  \subref{fig:primary_cell} shows a primary cell, and \subref{fig:secondary_cell} shows a secondary cell in a default patterning.}

%\caption{ \captionsetup{subrefformat=parens} \label{fig:PATTERN} \textit{In silico} experiments with parameter values $a=0.1$, $b=100$, $\mu=\rho=10$, $k=h=2$, $\gamma\rightarrow\infty$, $\nu=1$ and $k^{sp}=50$. \subref{fig:pat_hom}, \subref{fig:pat_9b8}, \subref{fig:pat_rand_p}, \subref{fig:pat_rand} and \subref{fig:10b8_p} show the steady state Notch patterning for differing initial conditions of Notch. \subref{fig:notch_hom}, \subref{fig:notch_9b8}, \subref{fig:notch_rand_p}, \subref{fig:notch_rand} and \subref{fig:10b8_n} show the associated time series solutions of intracellular Notch. Note \subref{fig:notch_9b8} includes inset of early time solutions to show initial heterogeneity. Videos of these simulations can be found in \hyperlink{SI_1}{SI Movie 1} -- \hyperlink{SI_5}{SI Movie 5}.  \subref{fig:primary_cell} shows a primary cell, and \subref{fig:secondary_cell} shows a secondary cell in a default patterning.}

\end{figure}

In their 1996 paper \cite{collier1996pattern}, \citeauthor{collier1996pattern} explained that a default patterning of one primary cell (low Notch level) for every two secondary cells (high Notch level) is the dominant ordering of cells at steady state, as shown in Figure \ref{fig:pat_9b8}. This ordering is such that the neighbouring set of each primary cell is exactly six secondary cells (see Figure \ref{fig:primary_cell}), while the neighbouring set of each secondary cell consists of three primary and three secondary cells in an alternating fashion (see Figure \ref{fig:secondary_cell}).\\

However, we see that the initial conditions of Delta and Notch influence the final pattern formed within the tissue (i.e. when the tissue is at steady state). Specifically, for homogeneous initial conditions of Delta and Notch, homogeneity is maintained as per Figure \ref{fig:pat_hom}. Time series solution are presented in Figure~\ref{fig:notch_hom}, which shows that the intracellular Notch levels across the tissue adjust to a homogeneous level (see \hyperlink{SI_1}{SI Movie 1} for a video of the simulation). However, if even a single cell is initialised away from this homogeneous value (tested down to variations of $10^{\,-16}$) then the default Notch patterning was obtained, see Figure \ref{fig:pat_9b8}. Moreover, looking at the time series solutions for the default patterned tissue (Figure \ref{fig:notch_9b8}), we can see that the patterning propagates from the initially seeded cell (see \hyperlink{SI_2}{SI Movie 2} for a video of the simulation).

It should be noted, that the system is quite sensitive to initial conditions of Delta and Notch. This can be seen by running two simulations from different initial random initial conditions, Figures \ref{fig:pat_rand_p} (\ref{fig:notch_rand_p}, see \hyperlink{SI_3}{SI Movie 3} for a video of the simulation)) and \ref{fig:pat_rand} (\ref{fig:notch_rand}, (see \hyperlink{SI_4}{SI Movie 4} for a video of the simulation)). The particular random conditions used in Figure~\ref{fig:pat_rand_p} are able to achieve the default Notch patterning, and we see from the time series solution Figure~\ref{fig:notch_rand_p} that the tissue is able to reorient itself and establish distinct patterning. However, those random initial conditions used within Figure~\ref{fig:pat_rand} are suitable, resulting instead in a partly patterned tissue, containing unresolved errors, as we see from the time series solution Figure~\ref{fig:notch_rand}, the tissue does attempt to achieve the default patterned state, but distinct (stable) errors occur (see \hyperlink{SI_5}{SI Movie 5} for a video of the simulation).

Patterning initiates from irregularities within the Delta and Notch levels of cells, this irregularity then propagates radially. If we were to consider infinite tissue structures, then the default patterning shown in Figure \ref{fig:pat_9b8} would emerge, referred to as a period 3 pattern \cite{collier1996pattern}. However, when considering periodic boundary conditions, which is often the case in \textit{in silico} experiments \cite{osborne2010hybrid, fletcher2013implementing}, combined with the understanding of how irregularities propagate, we find that the patterning present in Figure \ref{fig:pat_9b8} is not always possible. We see from simulations on different tissue geometries starting with a single seeded cell (the most stable initial condition for patterning) that patterning is not possible on domains of arbitrary size. For example,  Figure~\ref{fig:10b8_p} shows the pattern generated on a tissue of $10\times8$ cells ($C_{x}=10$ and $C_{y}=8$), which show errors in the patterning. The time series solutions, Figure~\ref{fig:10b8_n} shows that initially, the patterning  propagates from the initial irregularity. However, as this propagation meets up with itself, a discrepancy occurs which results in the error shown down the middle of the tissue.  

\begin{figure}[H]
\centering

\begin{subfigure}[b]{0.32\textwidth}
\centering\includegraphics[width=0.75\textwidth]{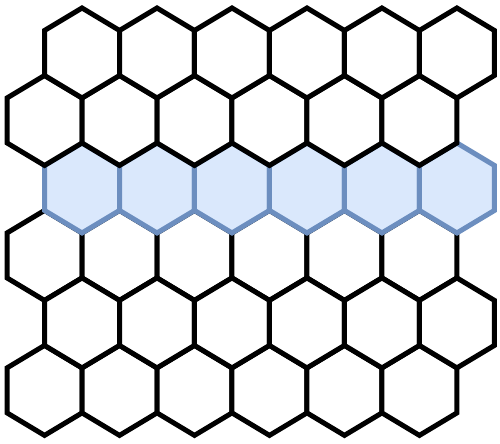}
\caption{\label{fig:trivial_knot}}
\end{subfigure}
\begin{subfigure}[b]{0.32\textwidth}
\centering\includegraphics[width=0.75\textwidth]{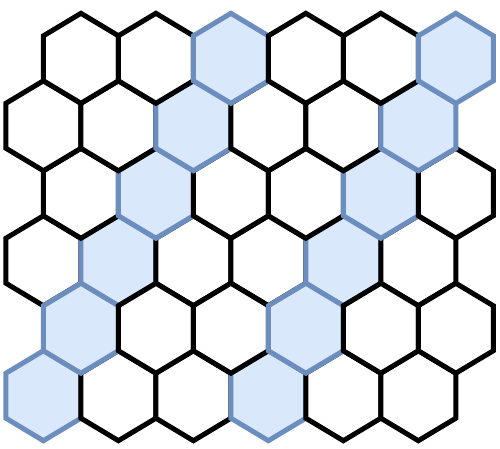}
\caption{\label{fig:non_trivial_knot}}
\end{subfigure}
\begin{subfigure}[b]{0.32\textwidth}
\centering\includegraphics[width=0.75\textwidth]{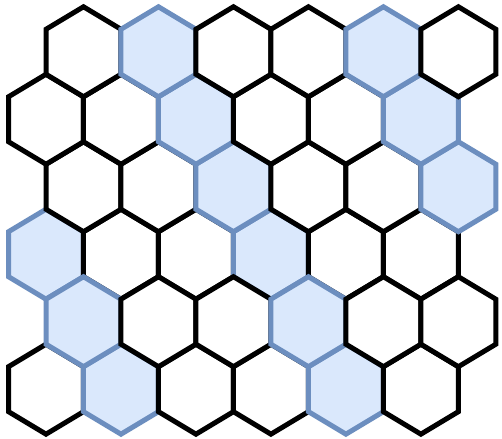}
\caption{\label{fig:non_trivial_knot_2}}
\end{subfigure}

\caption{ Examples of knots on a $6 \times 6$ size tissue. \subref{fig:trivial_knot} shows an example of a trivial knot. \subref{fig:non_trivial_knot} shows an example of a non-trivial knot. \subref{fig:non_trivial_knot_2} shows another example of a non-trivial knot. \label{fig:torus_knots} }
%\caption{\captionsetup{subrefformat=parens} Examples of knots on a $6 \times 6$ size tissue. \subref{fig:trivial_knot} shows an example of a trivial knot. \subref{fig:non_trivial_knot} shows an example of a non-trivial knot. \subref{fig:non_trivial_knot_2} shows another example of a non-trivial knot. \label{fig:torus_knots} }
\end{figure}

%In order to investigate the effect of geometry further we consider varying the tissue size.
To determine the suitable geometry size, $C_{x}$ and $C_{y}$, such that the tissue supports a default pattern, we first observe that the patterning is a period 3 pattern, meaning that if we start at a given cell, and we move away from it in a straight line, then as we move, we would notice that the pattern repeats itself every 3 cells. 
Due to the packing of hexagonal lattices with periodic boundary conditions, we require that the number of vertical cells be $C_{y} = 2n$, where $n \in \mathbb{N}$.
We note that there are three axes of symmetry on hexagonal lattices: the horizontal axis, another at $60^o$ to the horizontal, and another at $120^o$ to the horizontal. Due to the periodicity of the tissue, we now define what is called a torus knot \cite{livingston1993knot}, but on a lattice: if we start at a given location, and move along one of the axes of symmetry, we will always end up back where we started. 
For the tissue to support patterning, we require that the length of all torus knots along each of the 3 axes of symmetry be divisible by three. 
The easiest and most obvious torus knot is horizontal and is shown in Figure \ref{fig:trivial_knot}, which for a general tissue has size $C_{x} \times 2n$. Requiring that the length of this knot be divisible by 3, means we must have $C_{x} = 3m$, where $m \in \mathbb{N}$, resulting in tissue sizes of $3m \times 2n$. However, it is not immediately obvious that the length of the non-trivial knots, shown by Figures \ref{fig:non_trivial_knot} and \ref{fig:non_trivial_knot_2}, are also divisible by 3. The length of these knots can be written as:
\begin{align}
\text{knot length} & = \frac{\text{size of tissue}}{\text{number of disjoint knots}}.
\end{align}
For a tissue of $3m \times 2n$ cells, the size of the tissue is simply $6mn$. To find the number of disjoint knots, we first observe that for every $2n$ steps we traverse diagonally, we advance $n$ rows horizontally, due to periodicity. Therefore, the number of disjoint knots which can be supported on a tissue of size $3m \times 2n$ is going to be given by:
\begin{align}
\text{number of disjoint knots} = \gcd \left(3m, n\right),
\end{align}
which gives the length of the knots as:
\begin{align}
\text{knot length} & = \frac{6mn}{\gcd \left(3m, n\right)},
\end{align}
which is always divisible by 3.

Therefore, we can say that only tissues of sizes $3m \times 2n$ cells, where $m,n \in \mathbb{N}$ will support a default Notch pattern shown in Figure \ref{fig:pat_9b8}. Tissues of other sizes cannot support a default patterning, but rather a partly patterned tissue with errors, analogues to that shown by Figure \ref{fig:pat_rand}. The simulations presented in Figures~\ref{fig:pat_9b8} and \ref{fig:10b8_p} (along with other geometries, not shown for brevity) confirm this. We also note that in reality, biological tissue would not obey such a strict geometric constraint, nor would it exist on a toroidally connected domain. However, it is worth mentioning that the real tissue would be orders of magnitude larger than can feasibly be handled computationally, and so any errors observed in Notch patterning, such as those observed in Figures \ref{fig:10b8_p} and \ref{fig:10b8_n}, which are artifacts due to the domain specification would also be insignificant in proportion.

\subsection{Affinity Constants $a, b$ Control Cell Patterning}
\begin{figure}[H]
\centering
\begin{subfigure}[b]{0.49\textwidth}
\centering
\small \textbf{Steady State Notch with $a$, $b=100$}
\centering\includegraphics[width=\textwidth]{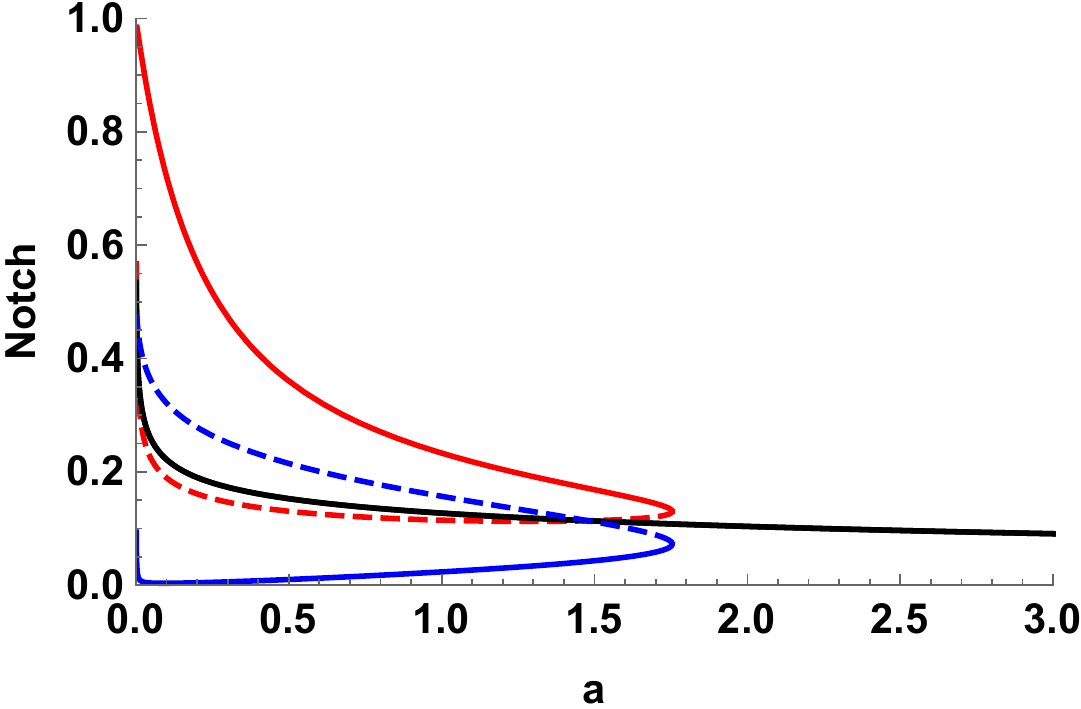}
\caption{\label{fig:ss_notch_w_a}}
\end{subfigure}
\begin{subfigure}[b]{0.49\textwidth}
\centering
\small \textbf{Steady State Notch with $b$, $a=0.01$}
\centering\includegraphics[width=\textwidth]{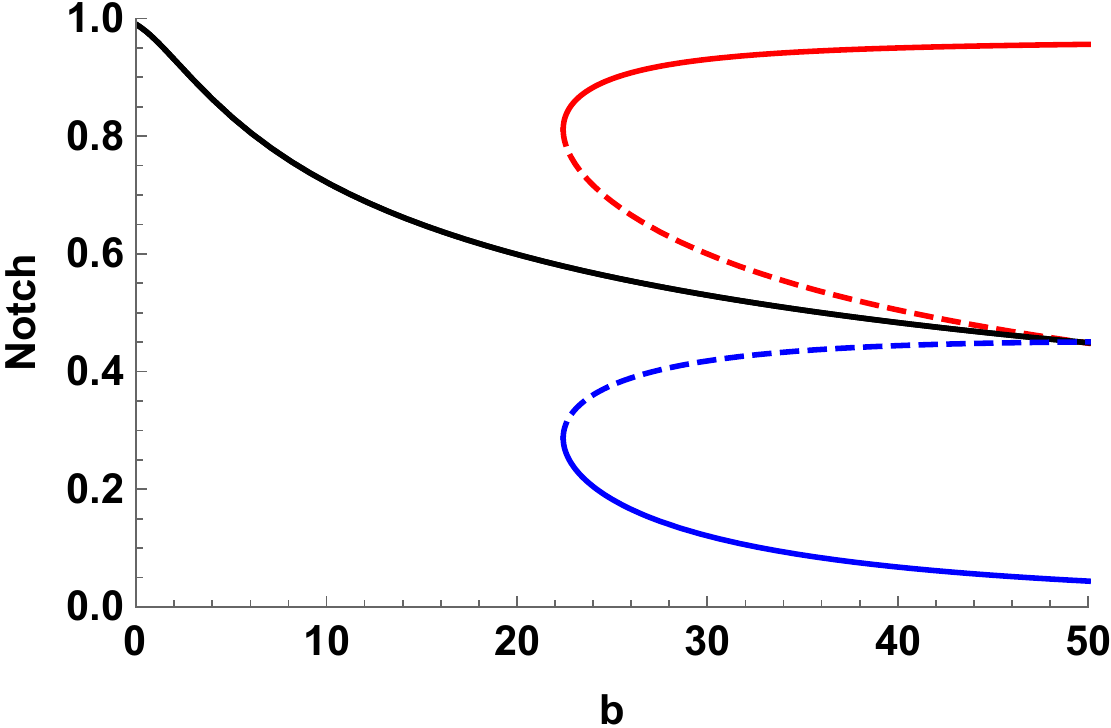}
\caption{\label{fig:ss_notch_w_b}}
\end{subfigure}

\begin{subfigure}[b]{0.8\textwidth}
\centering
\textbf{Relationship between $a_{\text{crit}}$ and $b_{\text{crit}}$}
\includegraphics[width=0.99\textwidth]{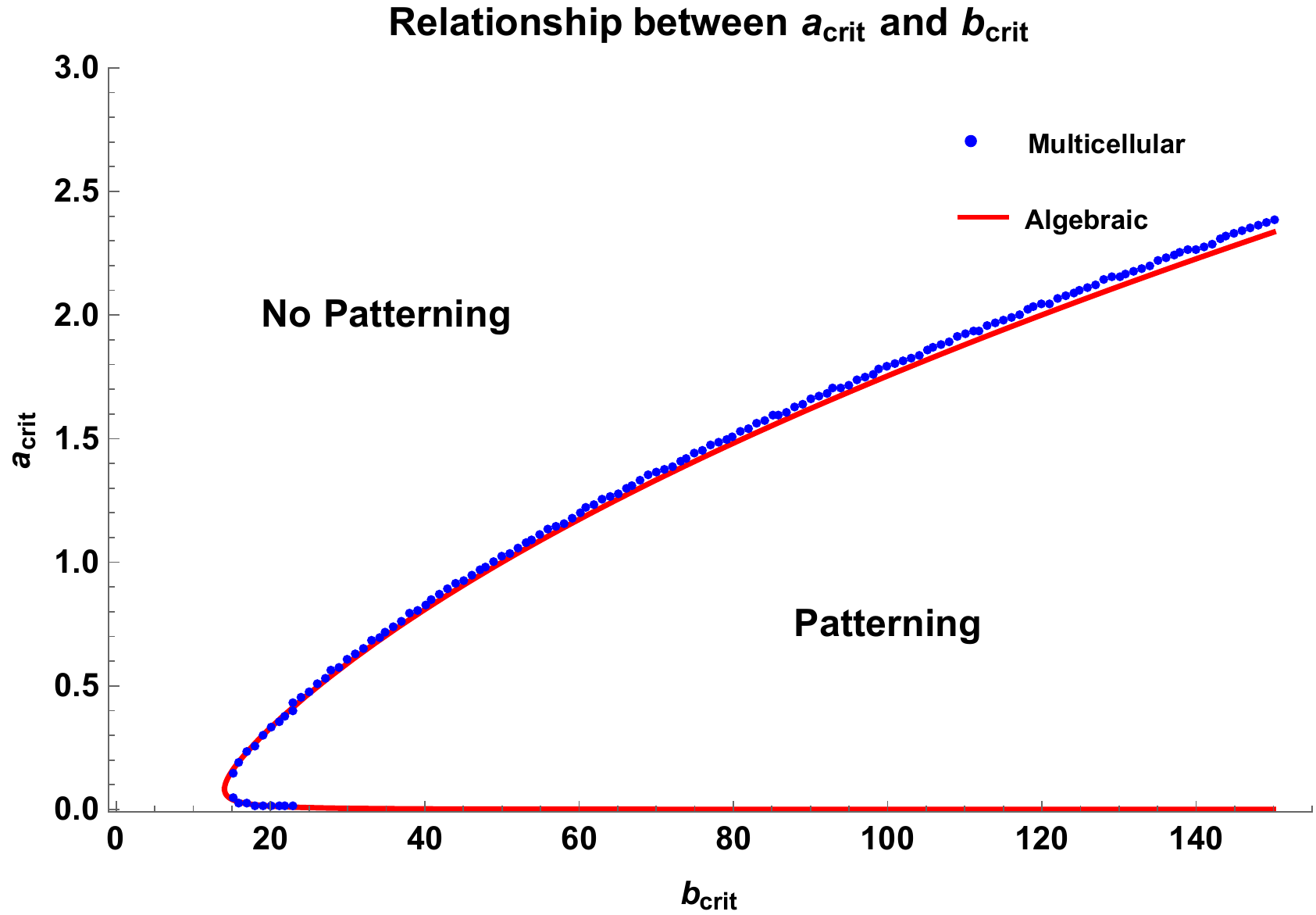}

\caption{\label{fig:an_ab_6_cells}}
\end{subfigure}

%\begin{figure}[H]
%\centering
%\begin{subfigure}{.015\textwidth}
%\centering {\rotatebox{90}{\hspace{6cm} \small Notch}
%\end{subfigure}
%\begin{subfigure}[b]{0.47\textwidth}
%\centering
%\small \textbf{Steady State Notch with $a$, $b=100$}
%\centering\includegraphics[width=\textwidth]{ss_notch_w_a_2}
%\text{$a$}
%\caption{\label{fig:ss_notch_w_a}}
%\end{subfigure}
%\begin{subfigure}{.015\textwidth}
%  \centering {\rotatebox{90}{\hspace{6cm} \small Notch}
%\end{subfigure}
%\begin{subfigure}[b]{0.47\textwidth}
%\centering
%\small \textbf{Steady State Notch with $b$, $a=0.01$}
%\centering\includegraphics[width=\textwidth]{ss_notch_w_b_2}
%\vspace{}\text{$b$}
%\caption{\label{fig:ss_notch_w_b}}
%\end{subfigure}
%
%\begin{subfigure}{.06\textwidth}
  %\centering {\rotatebox{90}{\hspace{8cm} \small \text{$a_{\text{crit}}$}}
%\end{subfigure}
%\begin{subfigure}[b]{0.8\textwidth}
%\centering
%\textbf{Relationship between $a_{\text{crit}}$ and %$b_{\text{crit}}$}
%\includegraphics[width=0.99\textwidth]{Robust_Pat_2}
%\small \text{$b_{\text{crit}}$}
%\caption{\label{fig:an_ab_6_cells}}
%\end{subfigure}
%\captionsetup{subrefformat=parens} 
\caption{ \subref{fig:ss_notch_w_a} and \subref{fig:ss_notch_w_b} shows the steady state Notch levels of the primary cells (blue), secondary cells (red) and the homogeneous level (black). Solid lines show the stable equilibrium and dashed lines show the unstable equilibrium, not numerically observed. \subref{fig:ss_notch_w_a} shows the Notch levels with varying affinity rate $a$, and \subref{fig:ss_notch_w_b} shows the Notch level with varying affinity rate $b$. \subref{fig:an_ab_6_cells} shows the relationship between the critical values $a_{\text{crit}}$ and $b_{\text{crit}}$ and the region within parameter space where patterning is permitted. The red solid line is results obtained from the simplified equations, and the blue markers are results obtained from multicellular simulations. Parameter values of $k=h=2$ and $\mu = \rho = 1$.\label{fig:ss_algebraic_eq} }
\end{figure}

In previous work, \citeauthor{collier1996pattern} stated that the affinity constants of both Notch and Delta control the existence of cell fate patterning, but did not state how. By further analysing the multicellular system, we are able to describe the effects that the affinity constants have on the existence of cell fate patterning, specifically we can say which parameter values allow patterning. By considering a static tissue which exhibits a default patterned state, such as that of Figure \ref{fig:pat_9b8}, we observe that there are at most two different cell configurations, up to rotation. The first is that of the primary fate cell, shown in Figure \ref{fig:primary_cell}, and the second is the secondary fate cell, shown in Figure \ref{fig:secondary_cell}. We therefore can write down the governing Notch and Delta equations of a primary fate cell in terms of $N_p$ and $D_p$, and similarly those for a secondary fate cell in terms of $N_s$ and $D_s$ as follows:
\begin{equation} \label{eq:original_DEs}
\begin{aligned}
\frac{d N_p}{dt} &= \mu \left( \frac{\bar{D}_p^k}{a+\bar{D}_p^k} - N_p\right), \quad & \frac{d N_s}{dt} =& \mu \left( \frac{\bar{D}_s^k}{a+\bar{D}_s^k} - N_s\right),\\
\frac{d D_p}{dt} &= \rho \left( \frac{1}{1+b N_p^h} - D_p\right),                        \quad  & \frac{d D_s}{dt} =& \rho \left( \frac{1}{1+b N_s^h} - D_s\right).
\end{aligned}
\end{equation}
We then make the observation from Figure \ref{fig:primary_cell} that the neighbours of all primary cells, consist of six secondary cells, to then write $\bar{D}_p$ as the following:
\begin{align}\label{Dp}
\bar{D}_p &= \frac{1}{6}\sum_{j \in M_p} D_j = \frac{1}{6}\left( 6 D_s\right) = D_s.
\end{align}
Similarly,  the neighbours of all secondary cells consists of three primary and three secondary cells, arranged in an alternating fashion, to then write $\bar{D}_s$ as the following:
\begin{align}\label{Ds}
\bar{D}_s &= \frac{1}{6}\sum_{j \in M_s} D_j = \frac{1}{6}\left( 3 D_p+3 D_s\right) = \frac{1}{2}\left( D_p+D_s\right).
\end{align}
%Substituting these forms of $\bar{D}_p$, Equation (\ref{Dp}), and $\bar{D}_s$, Equation (\ref{Ds}), into the differential Equation (\ref{eq:original_DEs}), leads to the simplified system:
Substituting  Equations (\ref{Dp}) and (\ref{Ds}), into Equation (\ref{eq:original_DEs}), leads to the simplified system:
\begin{equation} \label{eq:simplified_DEs}
\begin{aligned}
\frac{d N_p}{dt} &=\mu\left( \frac{{D_s}^k}{a+{D_s}^k} - N_p\right),  \,\, & \frac{d N_s}{dt} &= \mu \left( \frac{\left[ \frac{1}{2}\left(D_p+D_s\right) \right]^k}{a+\left[ \frac{1}{2}\left(D_p+D_s\right) \right]^k} - N_s\right), \\
\frac{d D_p}{dt} & = \rho \left( \frac{1}{1+b{N_p}^h} - D_p\right), \,\,  &\frac{d D_s}{dt} &= \rho \left( \frac{1}{1+b {N_s}^h} - D_s\right).
\end{aligned}
\end{equation}
Using the above set of equations, we were able to efficiently numerically determine how the affinity values affect the existence of cell fate patterning.
Steady states for Equation~(\ref{eq:simplified_DEs}) are shown, for varying affinity parameters, in Figures \ref{fig:ss_notch_w_a} and \ref{fig:ss_notch_w_b} . 
We define the critical values $a_{\text{crit}}$ and $b_{\text{crit}}$ to be the bifurcation points in $a$ and $b$ respectfully where Notch patterning ceases to exist. 
For example, considering Figure  \ref{fig:ss_notch_w_a}, for a fixed Delta affinity of $b=100$, the critical Notch affinity is $a_{\text{crit}} \approx 1.7538$ and considering Figure  \ref{fig:ss_notch_w_b}, for a fixed Notch affinity of $a=0.01$, the critical Delta affinity is $b_{\text{crit}} \approx 22.4304$.
The relationship between $a_{\text{crit}}$ and $b_{\text{crit}}$ can be found by solving the simplified system, given in Equation (\ref{eq:simplified_DEs}), invoking the steady state condition, and equating $N_p = N_s$ and $D_s = D_p$. 
The result is a polynomial in both the critical values $a_{\text{crit}}$ and $b_{\text{crit}}$, which can then be solved numerically. The relationship is shown in Figure \ref{fig:an_ab_6_cells} by the red solid line. The same relationship may also be found by solving the full multicellular system to steady state, and performing a parameter sweep to identify both $a_{\text{crit}}$ and $b_{\text{crit}}$, also shown in Figure \ref{fig:an_ab_6_cells} with blue markers.

From the results shown in Figure \ref{fig:ss_algebraic_eq} and the above discussion, we can therefore say that the affinity constants, $a$, which controls how readily Delta ligands bind to Notch receptors, and $b$, which controls how readily Delta is produced, completely govern the existence of Notch patterning on appropriate domains (i.e. domains where patterning is possible). When Delta-Notch binding events do not readily occur, or Delta is not easily expressed, then cells cannot differentiate.  

\subsection{Excessive Cell Turnover Inhibits Patterning}
\label{sec:dynamic}
We now consider how a finite cell turnover rate, $\gamma$, affects the patterning (and therefore differentiation) of the tissue in a dynamic steady state. We fix the biomechanical and biochemical parameter values of the model to those of Table \ref{table:parameter_values}. Note that the affinity parameters are chosen to allow patterning.

\begin{table}[h!]
\centering
\begin{tabular}{ |c|c|c| } 
 \hline
Parameter & Description & Value\\
 \hline
 $\mu$ & Rate of notch degradation & 10 \\ 
 $\rho$ & Rate of delta degradation & 10 \\ 
 $a$ & Notch affinity & 0.01 \\ 
 $b$ & Delta affinity & 100 \\ 
 $k$ & Notch synthesis exponent & 2 \\ 
 $h$ & Delta synthesis exponent & 2 \\ 
  \hline
 $\nu$ & Drag coefficient & 1 \\ 
 $k_{ij}^{\text{sp}}$ & Spring strength & 50 \\ 
 $\varepsilon$ & initial cell separation & 0.001  \\
  \hline
 $I_p$ & Notch primary cell threshold & 0.1 \\ 
 $I_s$ & Notch secondary cell threshold & 0.6 \\ 
 \hline
\end{tabular}
\caption{Biomechanical, biochemical and patterning proportion parameter values.}
\label{table:parameter_values}
\end{table}

To quantify the patterning throughout the tissue, we specify a Notch threshold for differentiated cells, $I_p$, for primary cells and $I_s$ for secondary cells, with $I_p < I_s$. We then categorise each of the cells within the tissue as either primary ($N_i \leq I_p$), secondary ($N_i \ge I_s$), or undifferentiated ($I_p < N_i < I_s $). These threshold values can be determined by solving Equation (\ref{eq:simplified_DEs}) to steady state for a specified set of parameters (see Table \ref{table:parameter_values}). One then chooses $I_p$ and $I_s$ suitably. For example, considering the biochemical parameter values of $a=0.01$, $b=100$, $k=h=2$, then the steady state Notch values for primary cells is $N_p \approx 0.0113$, and for secondary cells is $N_s \approx  0.9614$. We therefore choose the threshold values of $I_p = 0.1$ and $I_s=0.6$ for primary and secondary cells respectively.

\begin{figure}[H]
\centering
\begin{subfigure}{.08\textwidth}
%  \centering \rotatebox{90}{\textbf{$\gamma = 10$}}
  \centering \rotatebox{90}{\textbf{\hspace{0.75cm}$\gamma = 0.1$}}
\end{subfigure}
\begin{subfigure}{.20\textwidth}
  \centering
  \includegraphics[width=0.1175\textheight]{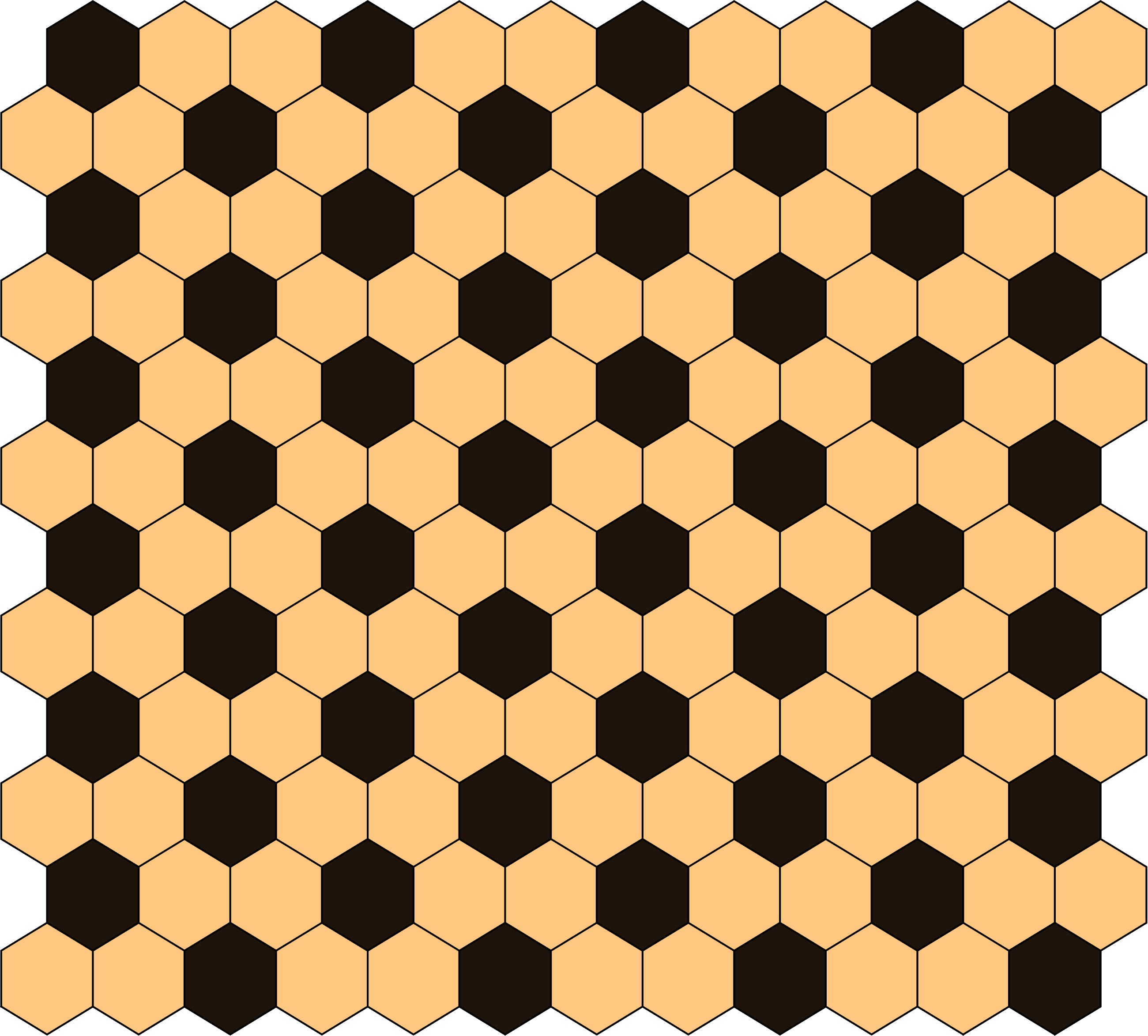}
  \caption{$t=0$hrs\label{fig:torus_evo_01_t0}}
\end{subfigure}%
\begin{subfigure}{.20\textwidth}
  \centering
  \includegraphics[width=0.125\textheight]{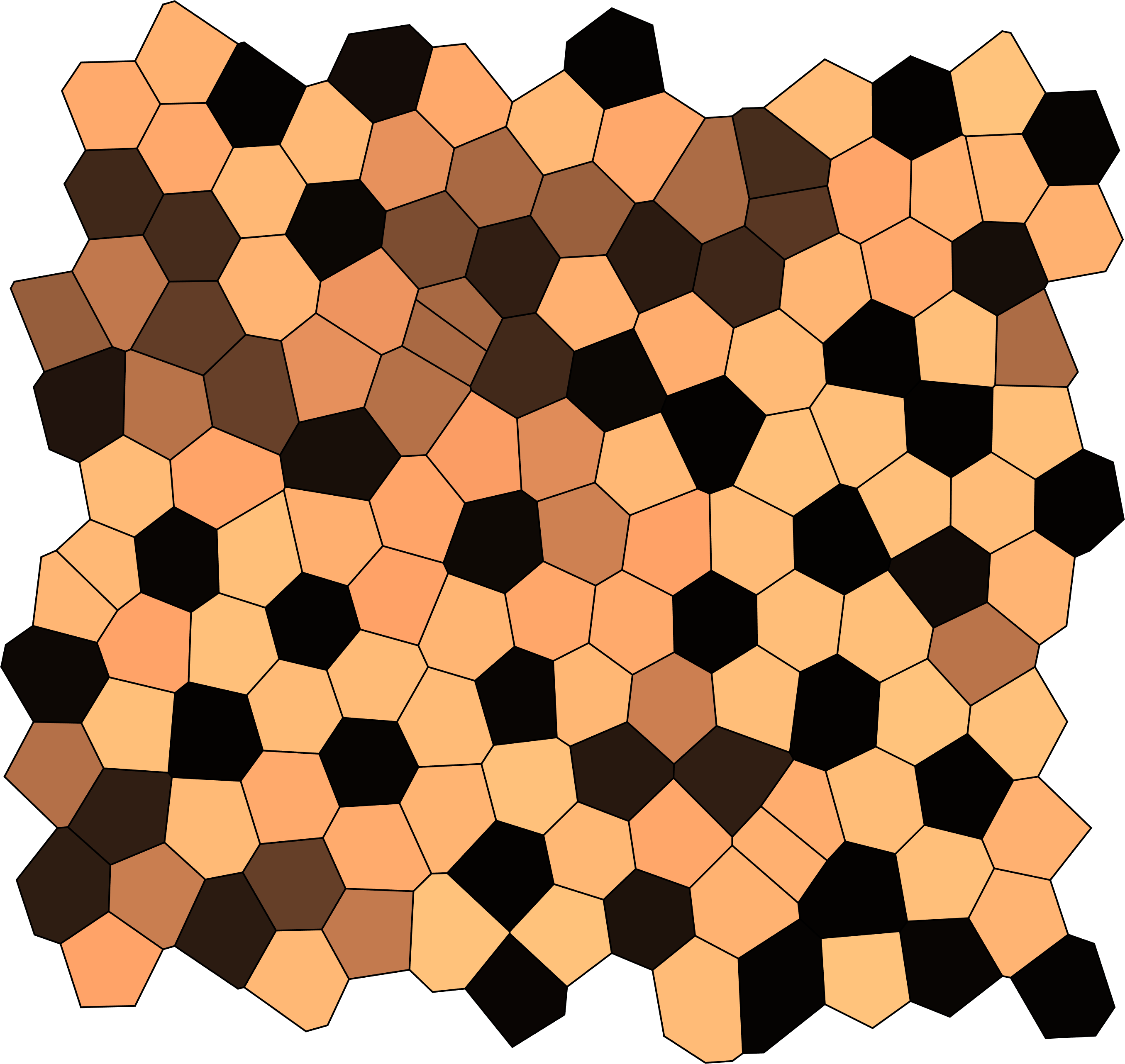}
  \caption{$t=40$hrs\label{fig:torus_evo_01_t40}}
  \label{fig:sub1}
\end{subfigure}
\begin{subfigure}{.20\textwidth}
  \centering
  \includegraphics[width=0.125\textheight]{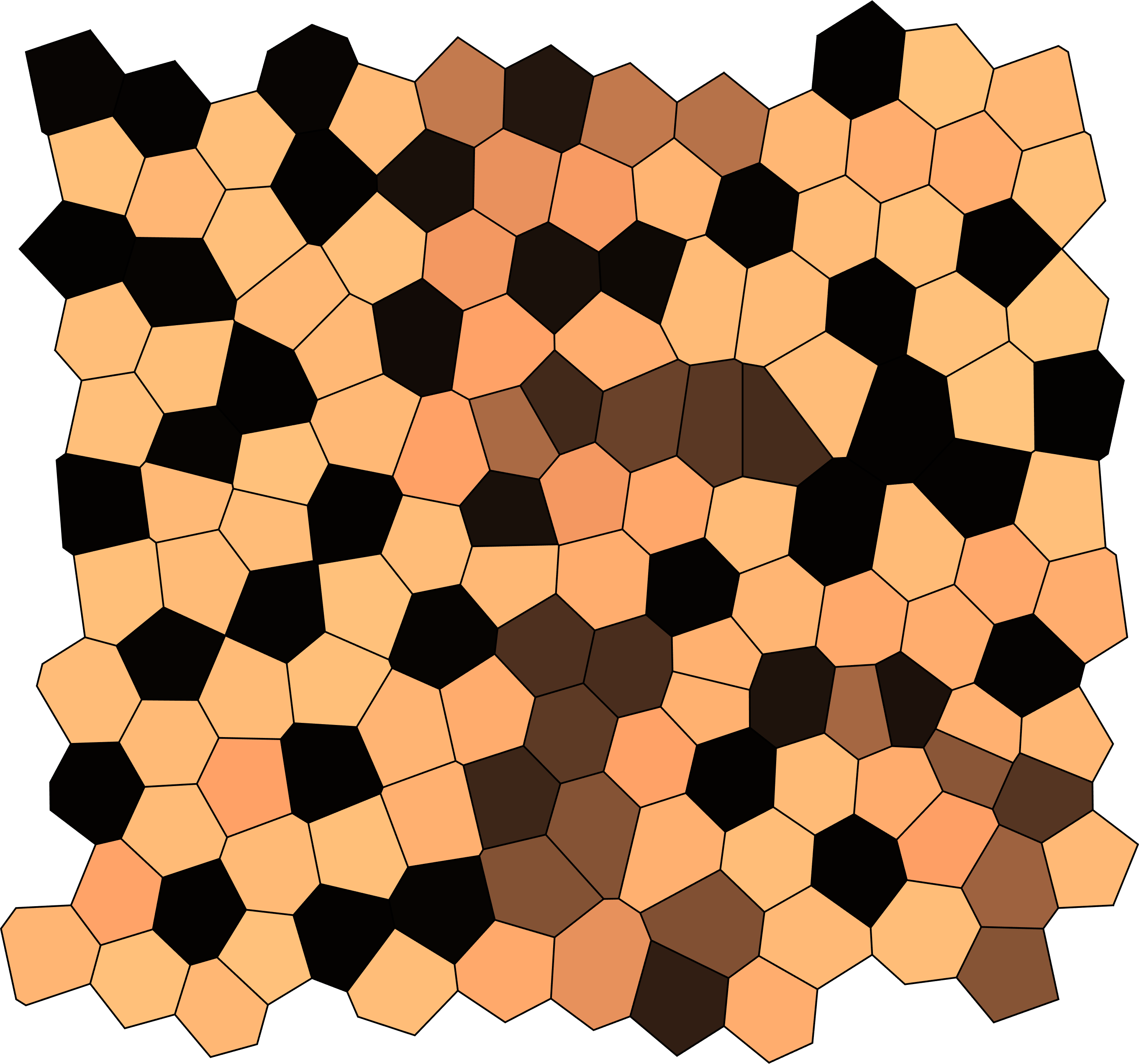}
  \caption{$t=80$hrs\label{fig:torus_evo_01_t80}}
\end{subfigure}
\begin{subfigure}{.20\textwidth}
  \centering
  \includegraphics[width=0.125\textheight]{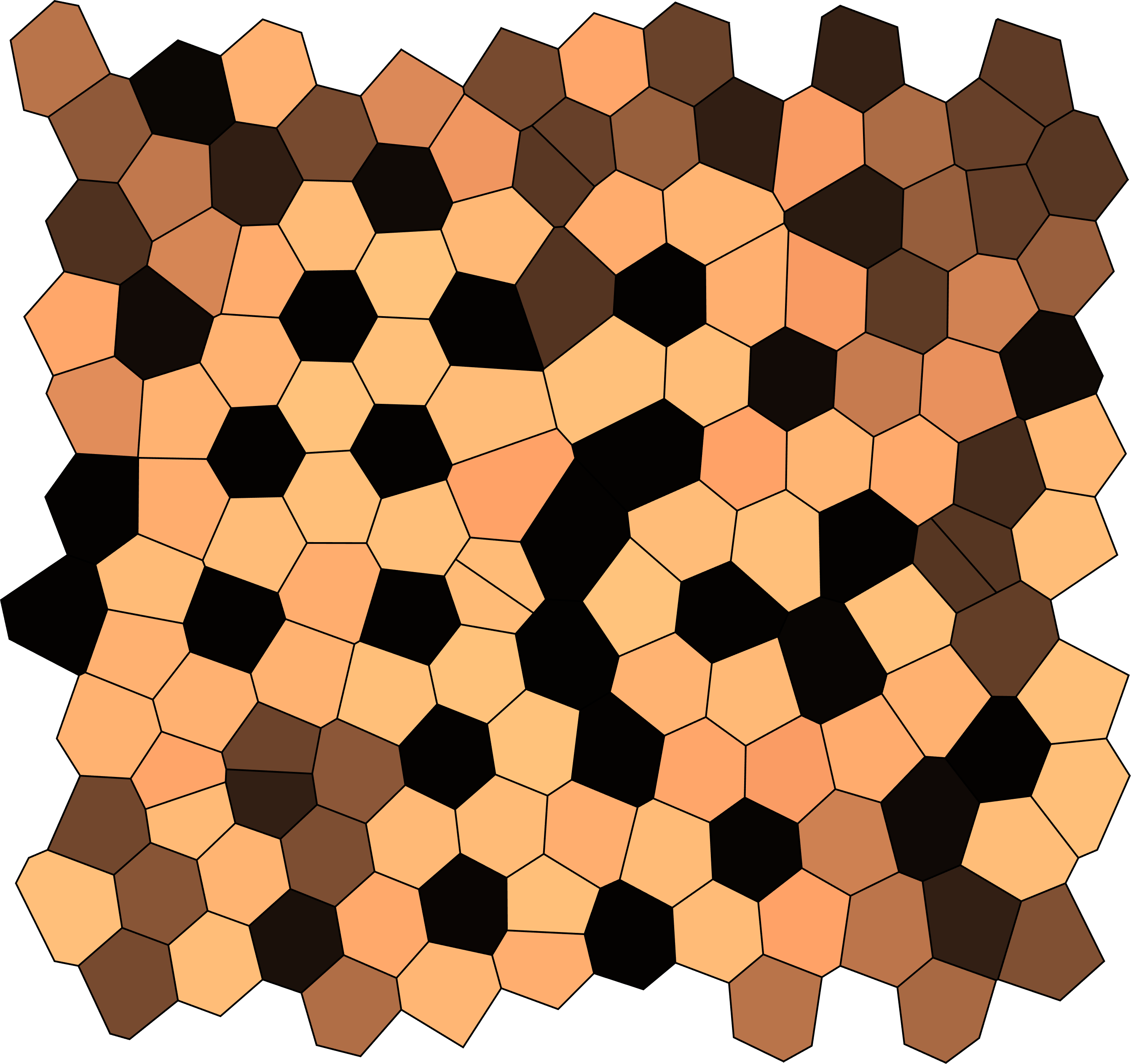}
  \caption{$t=120$hrs\label{fig:torus_evo_01_t120}}
\end{subfigure}
\begin{subfigure}{.08\textwidth}
  \centering
  \includegraphics[height=0.15\textheight]{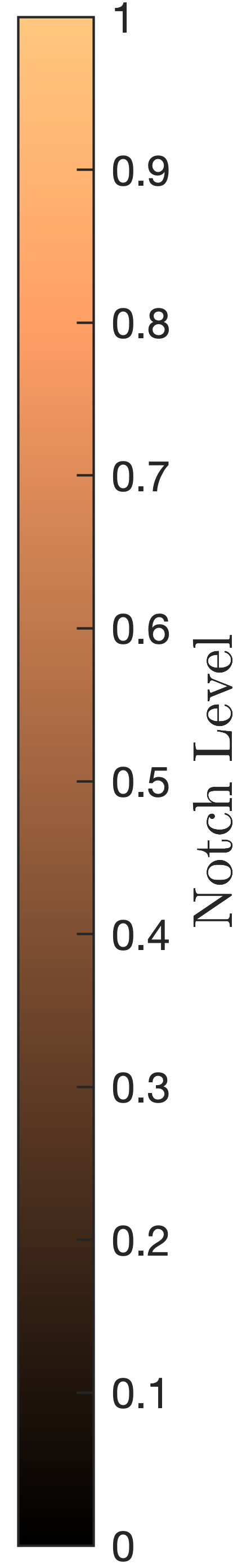}
  %\caption{$t=168$hrs\label{fig:colorbar}}
\end{subfigure}
\\
\vspace*{0.5cm}

\begin{subfigure}{.08\textwidth}
  \centering {\rotatebox{90}{\hspace{1cm}$\Omega$}}
\end{subfigure}
\begin{subfigure}{0.91\textwidth}
\centering
\textbf{Patterning of Tissue with Time}
  \centering
  \includegraphics[width=0.98\textwidth]{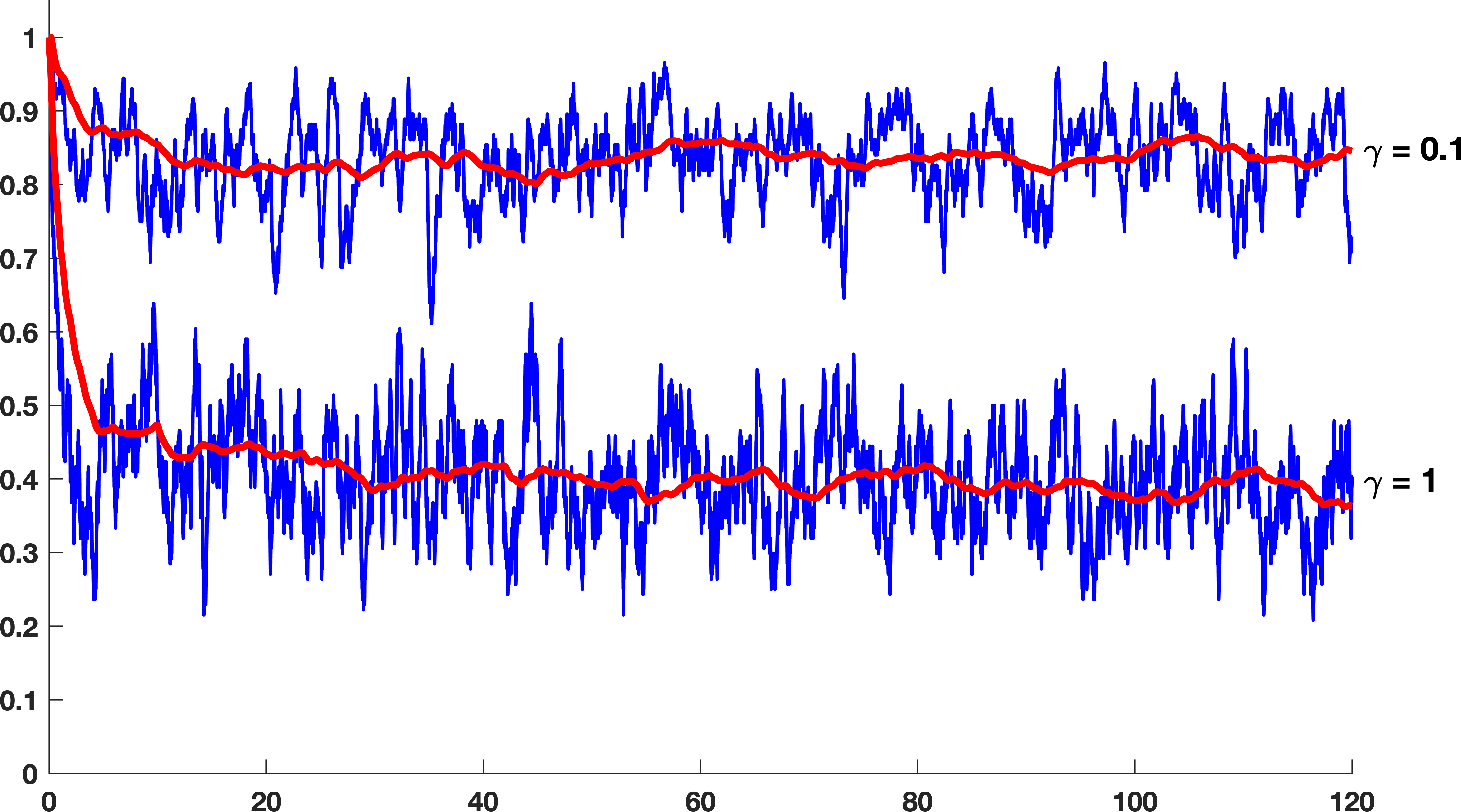}
  \text{time (hrs)}
  \caption{\label{fig:torus_patterning_evo_sub}}
\end{subfigure}
\\

\vspace*{0.5cm}
\begin{subfigure}{.08\textwidth}
%  \centering \rotatebox{90}{\textbf{$\gamma = 1$}}
  \centering \rotatebox{90}{\textbf{\hspace{0.75cm}$\gamma = 1$}}
\end{subfigure}
\begin{subfigure}{.2\textwidth}
  \centering
  \includegraphics[width=0.1175\textheight]{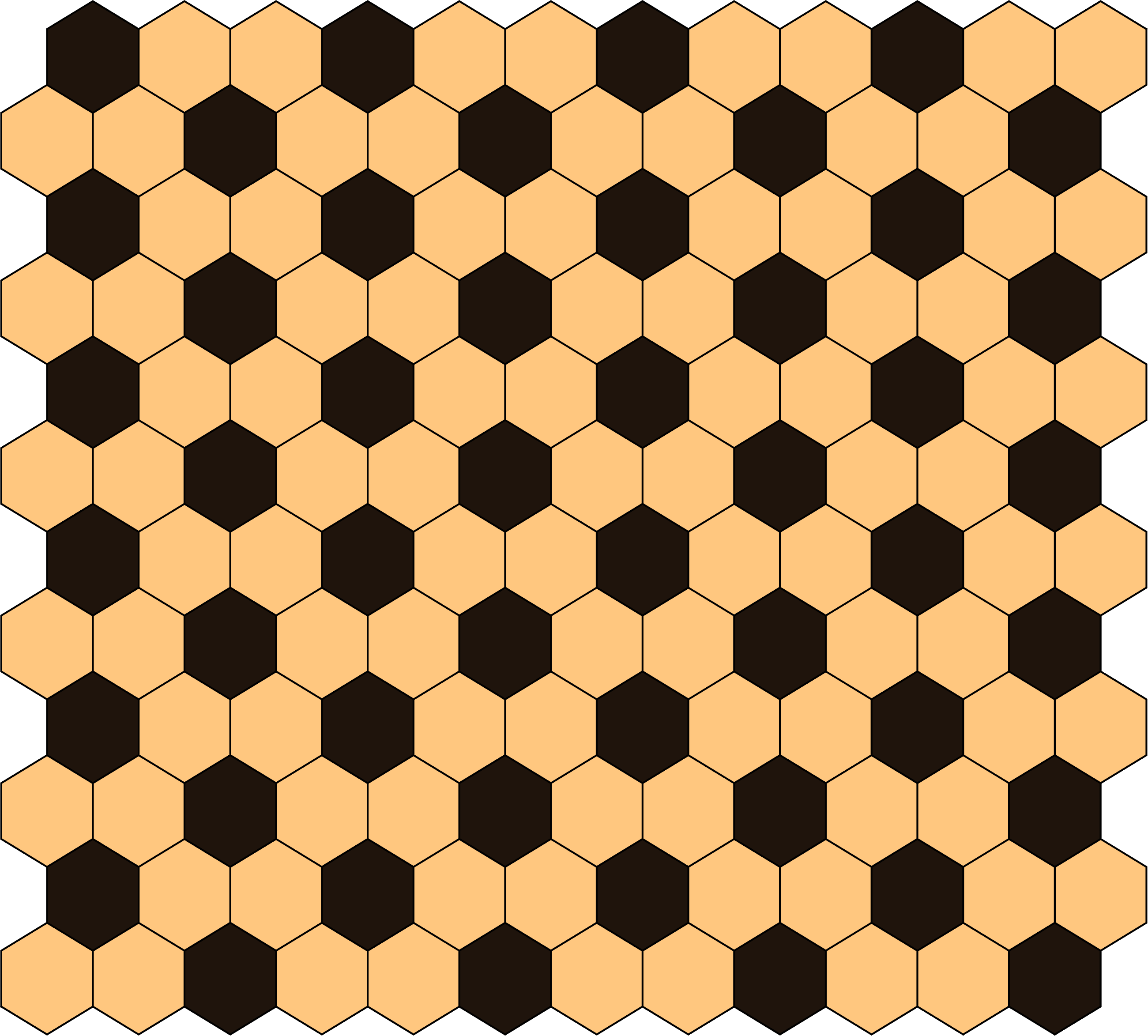}
  \caption{$t=0$hrs\label{fig:torus_evo_1_t0}}
\end{subfigure}%
\begin{subfigure}{.2\textwidth}
  \centering
  \includegraphics[width=0.125\textheight]{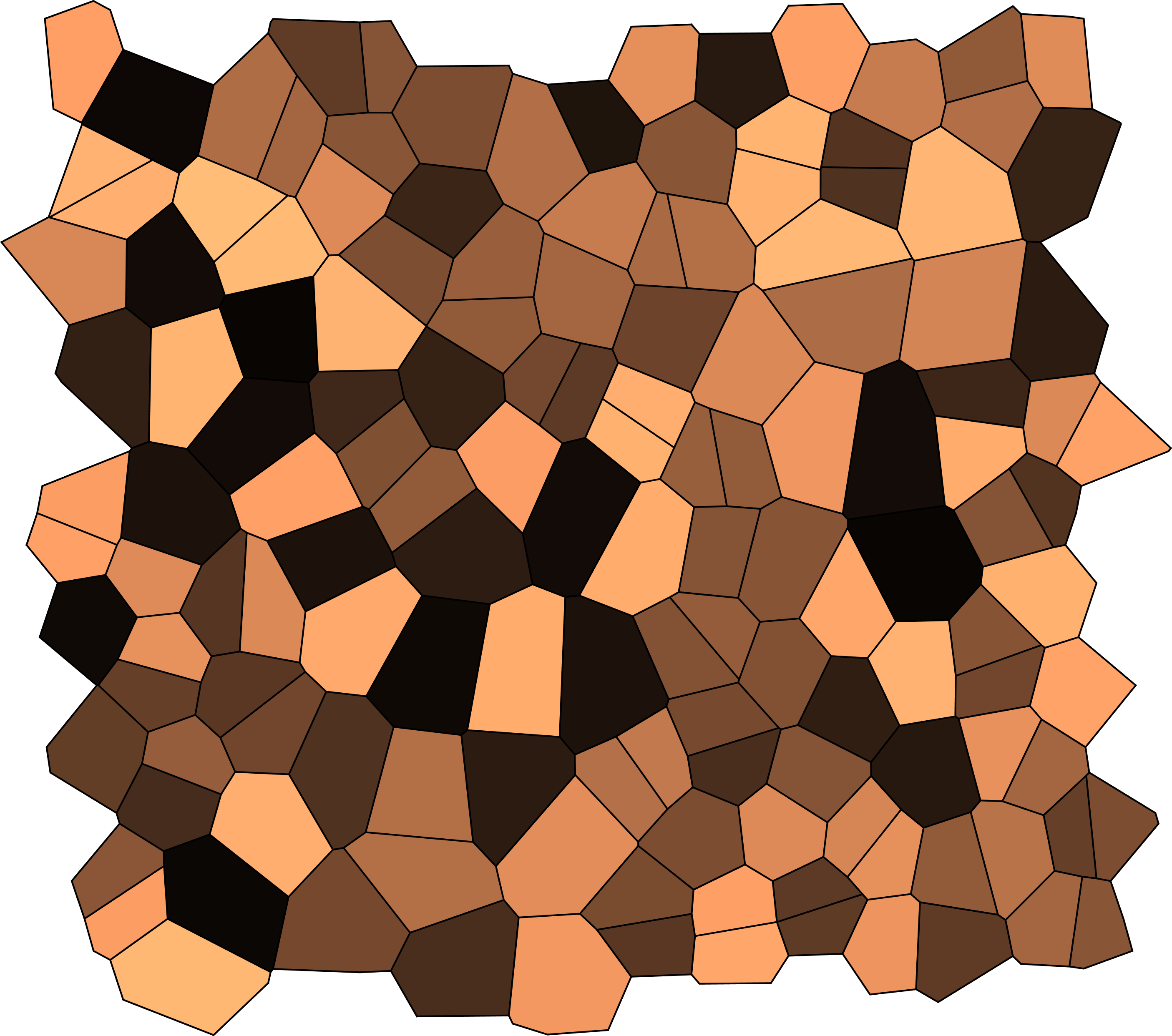}
  \caption{$t=40$hrs\label{fig:torus_evo_1_t40}}
  \label{fig:sub2}
\end{subfigure}
\begin{subfigure}{.2\textwidth}
  \centering
  \includegraphics[width=0.125\textheight]{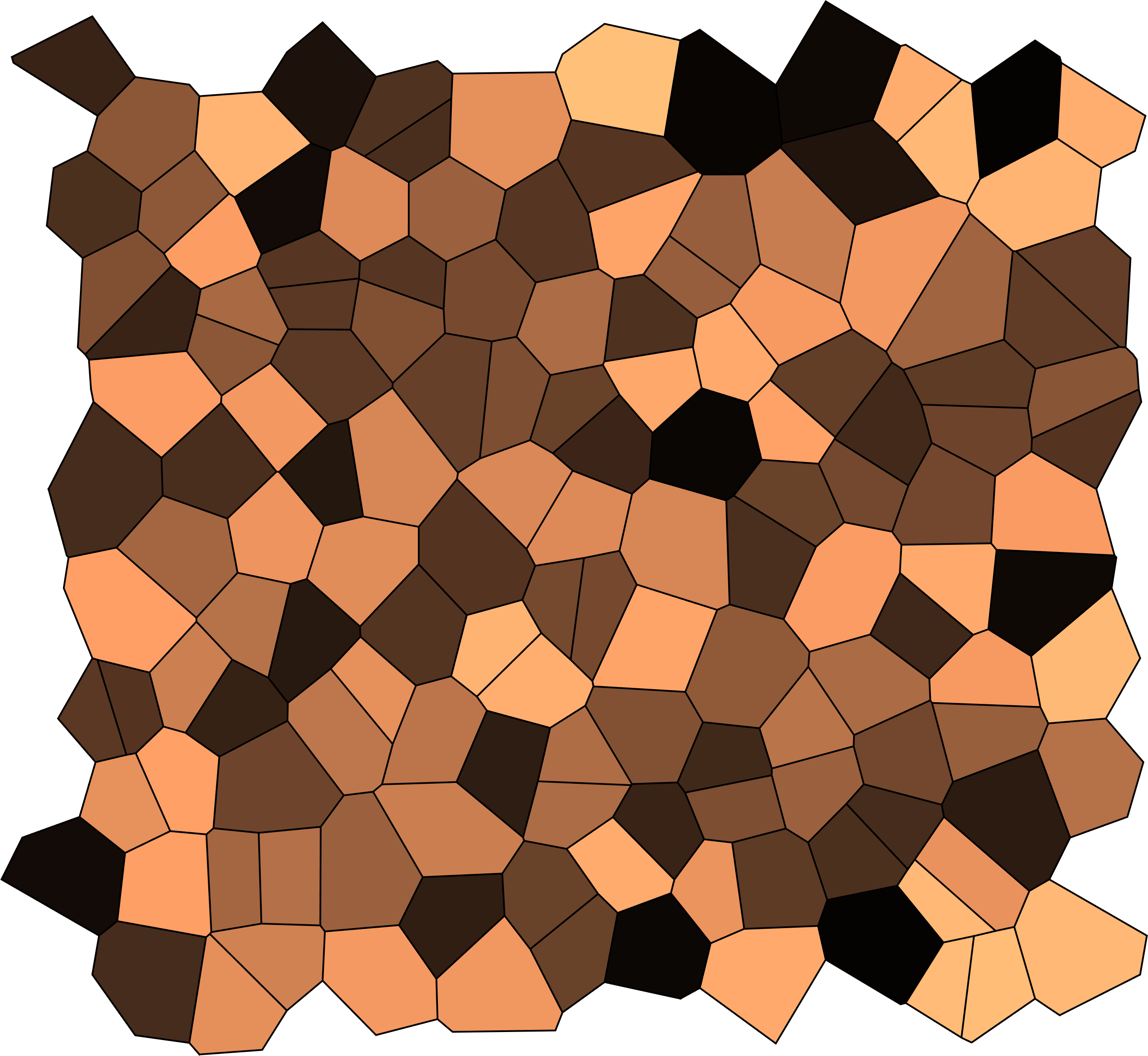}
  \caption{$t=80$hrs\label{fig:torus_evo_1_t80}}
\end{subfigure}
\begin{subfigure}{.2\textwidth}
  \centering
  \includegraphics[width=0.125\textheight]{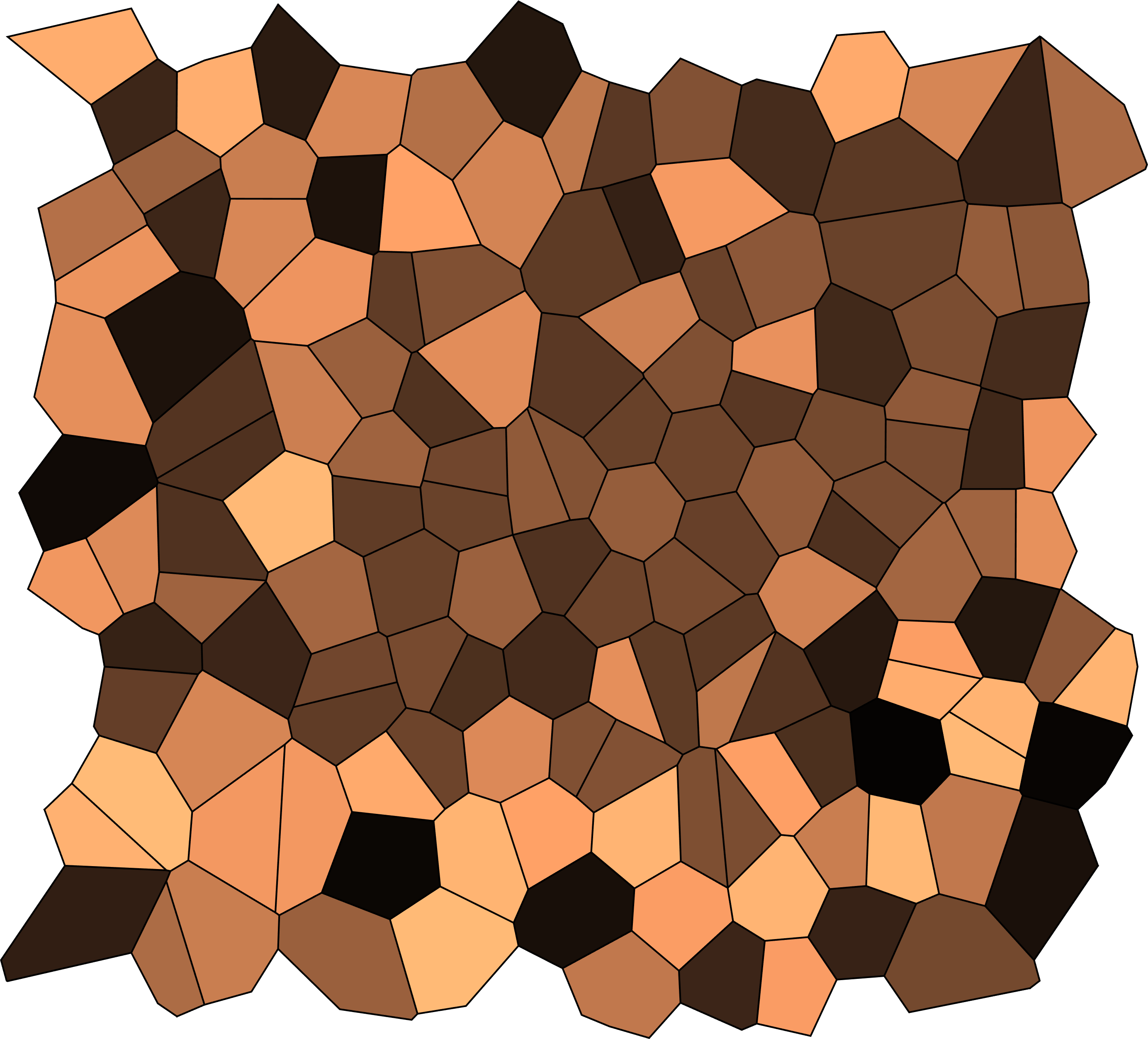}
  \caption{$t=120$hrs\label{fig:torus_evo_1_t120}}
\end{subfigure}
\begin{subfigure}{.08\textwidth}
  \centering
  \includegraphics[height=0.15\textheight]{cbar_f.png}
  %\caption{$t=168$hrs\label{fig:colorbar}}
\end{subfigure}
%\captionsetup{subrefformat=parens} 
\caption{ \label{fig:torus_patterning_evo} Figures \subref{fig:torus_evo_01_t0} - \subref{fig:torus_evo_01_t120} show a typical \textit{in silico} experiment of a toroidal tissue, with a cell turnover rate of $\gamma =0.1$ divisions per hour per cell (see SI Movie 6 for a video of the simulation). Figures \subref{fig:torus_evo_1_t0} - \subref{fig:torus_evo_1_t120} show a typical \textit{in silico} experiment of a toroidal tissue, with a cell turnover rate of $\gamma = 1$ division per hour per cell (see SI Movie 7 for a video of the simulation). Figure \subref{fig:torus_patterning_evo_sub} shows the corresponding evolution of the patterning for both $\gamma= 0.1$ (top) and $\gamma= 1$ (bottom), on a tissue size of $L_{x} = 12$cd and $L_y=6\sqrt{3}$cd.. The blue lines show the instantaneous patterning, while the red shows the moving average.}
\end{figure}

The proportion of the tissue which is in a patterned (differentiated) state, $\Omega(t)$, is then given by the total number of differentiated cells, divided by the total number of cells within the tissue at time $t$.
However, due to the dynamic nature of the system, we further consider a moving average of the instantaneous pattering, $\Omega(t)$, as:
\begin{align}
\bar{\Omega}\left(t; \delta \right) = \frac{1}{\delta}\int_{t-\delta}^{t} \Omega(\tau) d\tau .
\end{align}
Where $0<\delta<t$ is the length of the interval we sample. Figure \ref{fig:torus_patterning_evo} shows an example of two tissues evolving over time with different cell turnover rates of $\gamma = 0.1$ and $\gamma= 1$ respectively. Figure \ref{fig:torus_patterning_evo_sub} shows the instantaneous (blue) and average (red) proportion of  patterning throughout the tissue. We observe that the cell turnover rate, $\gamma$ influences the proportion of patterning within the tissue. Specifically larger levels of cell turnover result in less patterning (differentiation).

\begin{figure}[H]
\centering
 \textbf{Effects of Cell Turnover Rate on Patterning}
\includegraphics[width= 0.85\textwidth]{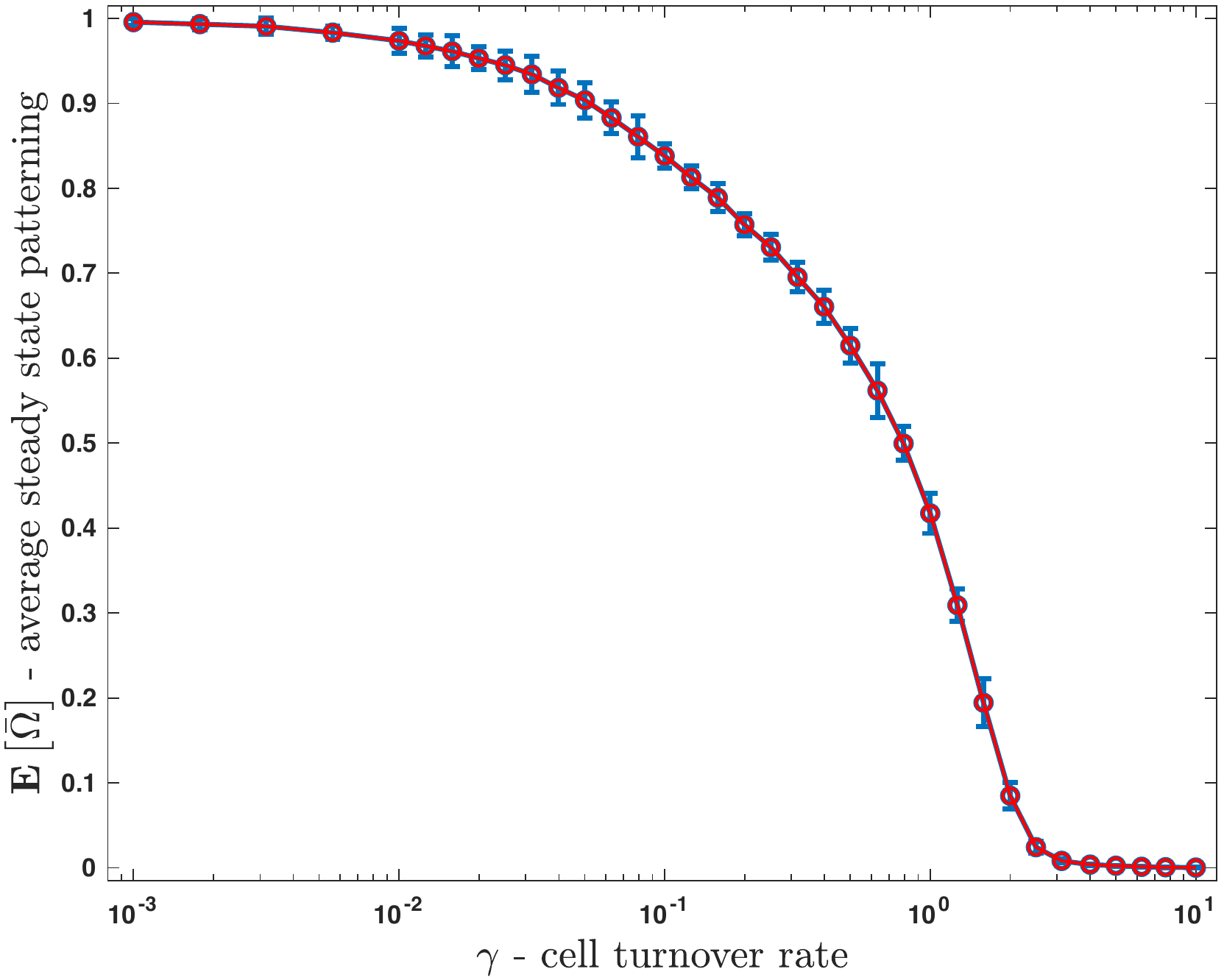}
\caption{ \label{fig:Torus_prol} Effects of cell turnover rate on patterning on a toroidal tissue, on a tissue size $L_{x} = 12$cd and $L_y=6\sqrt{3}$cd. Red markers show mean patterning, averaged over 20 \textit{in silico} experiments. Blue error bars represent the $95\%$ confidence interval.}
\end{figure}

To Further investigate the influence of the cell turnover rate on tissue patterning, we performed multiple \textit{in silico} experiments, with varying cell turnover rates of $\gamma \in \left[10^{\, -3} , 10^{\, 1}\right]$, results are shown in Figure \ref{fig:Torus_prol}, on a tissue size of $L_{x} = 12$cd and $L_y=6\sqrt{3}$cd. As the cell turnover rate, $\gamma$, is increased the Delta-Notch patterning is inhibited and cell differentiation suffers as a result. This is because as $\gamma$ increases, the neighbouring cells any given cell interacts with changes more rapidly, preventing the cell from achieving chemical equilibrium, which results in undifferentiated cells and hence inhibited Notch patterning. Specifically, as we increase the cell turnover rate from $\gamma  = 10^{\, -3}$, to $\gamma  = 0.1$, the tissue remains in a patterned state. Increasing the cell turnover rate from  $\gamma  = 0.1$ to $\gamma  = 2$, the tissue transitions from being in a patterned state to a homogeneous state, exhibiting no distinct cell fates. Further increases in $\gamma$ simply results in the homogeneous state.

\section{Discussion}
\label{sec:conclusion}
Throughout this paper, we have presented a mathematical model for cell fate selection on a general dynamic tissue. We couple the model of cell signalling via contact inhibition of \citeauthor{collier1996pattern} \cite{collier1996pattern}, with the model of cell dynamics of \citeauthor{meineke2001cell} \cite{meineke2001cell}.

On static tissues, we extended the work of \citeauthor{collier1996pattern} \cite{collier1996pattern} to observe that both the tissue geometry, and initial Delta-Notch conditions govern the steady state distribution within the tissue. Specifically, we see that both can disrupt regular patterning, and even inhibit patterning altogether. We further observed that the affinity constants, $a$ (affinity for Delta ligands to bind to Notch receptors) and $b$ (affinity for Delta expression) completely govern Notch patterning on appropriate domains. Through an exploration of the parameter space, we determined the critical thresholds, $a_{\text{crit}}$ and $b_{\text{crit}}$ which permit Notch patterning. These results suggest that when the Delta-Notch binding event does not occur readily, then cells do not become differentiated and cell fate selection does not occur. Similar results occur when Delta ligands are not readily produced.

Moreover, by observing a rotational symmetry of the steady state Notch patterning, we were able to deduce a reduced system of ordinary differential equations which is capable of capturing the steady state behaviour of the system, while significantly reducing the computational complexity in comparison to the full multicellular system. We were then able to find an algebraic expression relating the bifurcation parameters $a_{\text{crit}}$ and $b_{\text{crit}}$.\\

Lastly, we looked at how cell turnover rate affects Notch patterning within dynamic tissues. We found that a fast cell turnover rate inhibits Notch patterning, resulting in a homogeneous tissue. For cell turnover rates ranging from $\gamma =  0.1$ divisions per hour, to $\gamma = 2$ divisions per hour, the tissue transitions from being almost fully patterned to a homogeneous tissue, with no distinct Notch patterning. This loss of Notch patterning occurs as we increase the divisions per hour since in turn, the neighbouring set of cells each cell is biochemically interacting with changes more rapidly. As these neighbouring interactions occur over a shorter time, the cells are prevented from achieving chemical equilibrium. This lack of equilibrium is seen as undifferentiated cells. These results suggest that the cell turnover rate plays a crucial role in maintaining healthy tissues, with higher uncontrolled cell turnover rates leading to a unhealthy epithelium.\\

Future avenues of study include an analysis of how realistic tissue geometries, such as those of the colonic crypt, behave. Specifically, within the crypts, it is known that the level of Wnt signalling a cell receives governs cell proliferation within the crypt \cite{gregorieff2005wnt}. 
It is further known that Wnt signalling is known to be greatest at the base of the crypt, and decreases up the crypt axis \cite{gregorieff2005wnt}. 
One possible way to model this phenomenon in 2D would be to consider a cylindrical geometry, with a section near the base which is allowed to proliferate and cell sloughing near the top, similar to those presented in \cite{osborne2010hybrid, vanLeeuwen2009cell}. 
This geometry could then determine how cell differentiation evolves as cells migrate towards the top of the crypt.
There is also a known cross-talk between the Notch and the Wnt pathways within the cell \cite{kay2017role}. Including these dynamics would further allow us to a obtain full, realistic model of cellular signalling. 
Finally, the above model naturally resides on a 2D tissue. However, the development and dynamics of real biological systems is rarely captured with 2D projections, but rather reside as 2D surfaces in 3D space. Thus the development of a 3D model which is capable of describing realistic tissue deformations is needed \cite{dunn2013computational}.

\section*{Acknowledgements}

This research was supported by an Australian Government Research Training Program (RTP) Scholarship (awarded to DPJG).

\section*{Supplementary Information}

\begin{description}
 \item[\hypertarget{SI_1}{SI Movie 1.}] {Video of simulation from Figures \ref{fig:pat_hom} and \ref{fig:notch_hom}, which shows how homogeneous Notch initial conditions lead to a homogeneous tissue.\\
 \texttt{SI\_Movie\_1.mp4}}
 \item[\hypertarget{SI_2}{SI Movie 2.}] {Video of simulation from Figures \ref{fig:pat_9b8} and \ref{fig:notch_9b8}, which shows how a perturbation away from homogeneous Notch initial conditions lead to a patterned tissue. This also shows how the patterning radially propagates from the perturbed cell.\\
 \texttt{SI\_Movie\_2.mp4}}
 \item[\hypertarget{SI_3}{SI Movie 3.}] {Video of simulation from Figures \ref{fig:pat_rand_p} and \ref{fig:notch_rand_p}, which shows random initial Notch conditions may lead to a patterned tissue.\\
 \texttt{SI\_Movie\_3.mp4}}
\item[\hypertarget{SI_4}{SI Movie 4.}] {Video of simulation from Figures \ref{fig:pat_rand_p} and \ref{fig:notch_rand_p}, which shows random initial Notch conditions may lead to only a partially a patterned tissue. This shows how Notch patterning is sensitive to initial conditions.\\
 \texttt{SI\_Movie\_4.mp4}}
\item[\hypertarget{SI_5}{SI Movie 5.}] {Video of simulation from Figures \ref{fig:pat_rand_p} and \ref{fig:notch_rand_p}, which shows how the tissue size effects the final patterned tissue. Here, the tissue size is not suitable, resulting in a mismatch in patterning as the patterning propagation meets up with itself, due to periodicity.\\
 \texttt{SI\_Movie\_5.mp4}}
\item[\hypertarget{SI_6}{SI Movie 6.}] {Video of simulation from Figures \ref{fig:torus_evo_01_t0} -- \ref{fig:torus_evo_01_t120}, which shows a dynamic tissue evolving over time, with a cell turnover rate of $\gamma = 0.1$, from times $t=0$hrs to $t=20$hrs. Here, we can see that the tissue reaches mostly patterned state.\\
 \texttt{SI\_Movie\_6.mp4}}
\item[\hypertarget{SI_7}{SI Movie 7.}] {Video of simulation from Figures \ref{fig:torus_evo_1_t0} -- \ref{fig:torus_evo_1_t120}, which shows a dynamic tissue evolving over time, with a cell turnover rate of $\gamma = 1$, from times $t=0$hrs to $t=20$hrs. Here, we can see that the tissue reaches a partially patterned state, as the cell do not sufficiently inhibit their neighbours.\\
 \texttt{SI\_Movie\_7.mp4}}
\end{description}

\bibliography{bibAMathematicalModelOfCellFateSelectionOnADynamicTissue}

\end{document}